\documentclass[aip,jcp,preprint,superscriptaddress,floatfix]{revtex4-1}

\usepackage[utf8]{inputenc}
\usepackage{color}
\usepackage{graphicx}
\usepackage{mathtools}

\usepackage{tikz}
\usepackage{listings}
\usepackage{pgfplots}
\usepackage{pgfplotstable}
\usepackage[version=4]{mhchem}
\usepackage{threeparttable}

\usetikzlibrary{positioning,fit,calc}
\tikzstyle{flowbox} = [rectangle,minimum width=2.5cm, minimum height=1cm,text centered, draw=black]
\tikzstyle{superbox} = [rectangle,text centered, draw=black,dashed]
\tikzstyle{arrow} = [thick,->]
\tikzstyle{dash_arrow} = [thick,dashed,->]
\pgfplotsset{compat=1.3}

\newcommand{\beq}{\begin{equation}}
\newcommand{\eeq}{\end{equation}}
\newcommand{\beqa}{\begin{eqnarray}}
\newcommand{\eeqa}{\end{eqnarray}}
\newcommand{\mea}{\textit{et al.}}

\newcommand{\ket}[1]{\left|#1\right>}
\newcommand{\bra}[1]{\left<#1\right|}
\newcommand{\bPsi}{\bra{\Psi}}
\newcommand{\kPsi}{\ket{\Psi}}								
\newcommand{\kPhi}{\ket{\Phi}}						
\newcommand{\ksig}[1][]{\ket{k_{#1}}}			
\newcommand{\allStates}{k_1,\ldots,k_L}	
\newcommand{\allStatesopp}{k_1^{\prime},\ldots,k_L^{\prime}} 
\newcommand{\alldim}{a_1,\ldots,a_{L-1}}                     
\newcommand{\alldimop}{b_1,\ldots,b_{L-1}}                     
\newcommand{\ONvec}{\ket{k_1 \ldots k_L}}	
\newcommand{\ONvecb}{\bra{k_1^{\prime} \ldots k_L^{\prime}}}    
\newcommand{\ONstring}{k_1 \ldots k_L}		

\newcommand{\qcm}{\textsc{QCMaquis}}

\newcommand{\dir}{{\textsc{DIRAC}}}

\newcommand{\mmod}{\texttt{M8mod}}
\newcommand{\dyc}{[Dy(III)(\mmod)]$^{-1}$}

\begin{document}

\title{An efficient relativistic density-matrix renormalization group implementation in a matrix-product formulation}

\author{Stefano Battaglia}
\affiliation{ETH Z\"urich, Laboratorium f{\"ur} Physikalische Chemie, Vladimir-Prelog-Weg 2, 8093 Z\"urich, Switzerland}

\author{Sebastian Keller}
\affiliation{ETH Z\"urich, Laboratorium f{\"ur} Physikalische Chemie, Vladimir-Prelog-Weg 2, 8093 Z\"urich, Switzerland}

\author{Stefan Knecht}
\email{stefan.knecht@phys.chem.ethz.ch}
\affiliation{ETH Z\"urich, Laboratorium f{\"ur} Physikalische Chemie, Vladimir-Prelog-Weg 2, 8093 Z\"urich, Switzerland}

\begin{abstract}
We present an implementation of the relativistic quantum-chemical density matrix 
renormalization group (DMRG) approach 
based on a matrix-product formalism. 
Our approach allows us to optimize matrix product state (MPS) wave functions including a variational description 
of scalar-relativistic effects and spin–orbit coupling from which we can calculate, for example, first-order 
electric and magnetic properties in a relativistic framework. While complementing our 
pilot implementation (S. Knecht \textit{et al.}, \textit{J.~Chem.~Phys.}, \textbf{140}, 041101 (2014)) 
this work exploits all features provided by its underlying non-relativistic DMRG implementation based on an matrix product state 
and operator formalism. 
We illustrate the capabilities of our relativistic DMRG approach by studying the ground-state magnetization as well as  
current density of a paramagnetic $f^9$ dysprosium complex as a function of the active orbital space employed 
in the MPS wave function optimization. 
\end{abstract}

\maketitle

\section{Introduction}\label{Introduction}

With the advent of the density-matrix renormalization group (DMRG) approach \cite{whit92,whit93,scho05,scho11} in (non-relativistic) quantum chemistry
\cite{lege08,chan08,chan09,mart10,mart11,chan11a,wout14,kura14a,yana15,oliv15a,szal15,knec16a,chan16},  
approximate solutions for a complete-active-space (CAS)-type problem within chemical accuracy became 
computationally feasible on a routine basis albeit the fact that variational optimization  
merely involves a polynomial number of parameters. The combination with a self-consistent-field orbital optimization \textit{ansatz} (DMRG-SCF) \cite{zgid08,ghos08,zgid08b,sunq17b,mayi17b} makes active orbital spaces accessible that surpass the CASSCF limit by about five to six times. Yet, the construction of a (chemically) meaningful active orbital space poses a fundamental challenge that calls for automation  \cite{stei16a,stei16b,stei17a,sayf17a}.
In contrast to early work on \textit{ab initio} DMRG \cite{scho05}, recent efforts on the quantum-chemical DMRG algorithm \cite{scho11} focused on a variational optimization of a special class of \textit{ansatz} states called matrix product states (MPSs) \cite{ostl95,vers04b} in connection with a 
matrix-product operator (MPO) representation of the quantum-chemical Hamiltonian \cite{murg10a,kell15a,kell16,chan16,nakaxxx}. 

Combining the principles of quantum mechanics \cite{dira48} and special relativity \cite{eins1905} into relativistic quantum mechanics, Dirac \cite{dira28} put forward a (one-electron) theory which constitutes the central element of relativistic quantum chemistry \cite{dyal07,reih09}. In this framework, deviations in the description 
of the electron dynamics increase compared to (nonrelativistic) Schr{\"o}dinger quantum mechanics 
as one considers electrons moving (in the vicinity of a heavy nucleus) at velocities close to the speed of light.  
Consequently, the differences between results of relativistic quantum calculations 
employing a finite and an infinite speed of light, respectively --- which are for the latter case equivalent to results from 
conventional nonrelativistic calculations based on the Schr{\"o}dinger equation --- can serve as a definition for the chemical concept of ``relativistic effects". 

The latter are commonly split into kinematic relativistic effects (sometimes also referred to as scalar-relativistic effects) and 
magnetic effects both of which grow approximately quadratic with the atomic number $Z$ \cite{pyyk88}. 
A scalar-relativistic theory therefore considers only kinematic relativistic effects whereas a fully relativistic theory includes both kinematic relativistic and magnetic effects. 
Kinematic relativistic effects reflect changes of the non-relativistic kinetic energy operator while 
spin-orbit (SO) interaction, which is the dominant magnetic effect (for heavy elements), originates from a coupling of the electron spin 
to the induced magnetic field resulting from its orbital motion in the field created by  
the other charged particles, namely the nuclei and remaining electrons \cite{saue11a}. 
Encounters of relativistic effects are ubiquitous in the chemistry and physics of heavy elements compounds, see for example Refs.~\citenum{dyal07}, \citenum{reih09}\ and \citenum{auts12}\ and references therein. 
Besides kinematic relativistic effects, SO interactions are also expected to strongly influence the chemical bonding 
and reactivity in heavy-element containing molecular systems \cite{cott06,schw14}. For example, 
Ruud and co-workers \cite{demi16}\ as well as 
Gaggioli \textit{et al.} \cite{gagg16} recently highlighted distinguished cases of mercury- and gold-catalyzed reactions 
where SO coupling effects are driving the main reaction mechanisms by 
paving the way for catalytic pathways that would otherwise not even have been 
accessible. 

Considering a prototypical molecular, open-shell lanthanide complex, we aim in this work at 
the calculation of (static) magnetic properties which either require SO-coupled wave functions 
such as molecular g-factors and electron-nucleus hyperfine coupling, 
which are central parameters in electron paramagnetic resonance (EPR) spectroscopy \cite{harr78,kaup04}, 
or assume a particular simple form in a relativistic framework, including, for example, 
the magnetization and current density \cite{reih09}. We will pay particular attention to the latter two properties since 
they play a particular role for the eligibility of open-shell lanthanide tags in spin-labeling of protein complexes \cite{jesc07,clor09a,jesc12,gold14,bert17}. 

As open-shell electronic structures are often governed by strong electron correlation effects, 
their computational description calls for a multiconfigurational \textit{ansatz} \cite{szal12,roca12}\ which usually set out from a spin-free (non-relativistic or scalar-relativistic) formulation. Here, electron correlation is commonly split into a static contribution and a dynamic contribution with the latter often 
being treated in a subsequent step by (internally-contracted) multi-reference perturbation or configuration 
interaction theories \cite{szal12,roca12}. Their zeroth-order wave function is typically of CAS-type such as CASSCF or DMRG(-SCF) which are assumed to be able to adequately grasp static electron correlation effects.  
In this context, additional challenges may be encountered in particular for heavy-element complexes which arise from 
the large number of (unpaired) valence electrons to be 
correlated, i.e., the $(n-2)f\ (n-1)d\ ns\ np$\ manifold of 
$f$-elements, as well as the occurrence of near-degeneracies of 
electronic states. 
To arrive at SO-coupled eigenstates in such a spin-free setting, requires then to invoke a so-called two-step SO procedure 
(see for example 
Refs.~\citenum{malm02,roem15,sayf16,knec16b,muss17}\ for  
recent developments based on CAS- and DMRG-like  
formulations). Notably, this procedure relies on 
an additivity of electron correlation and SO effects or a weak 
polarization of orbitals due to SO interaction, or both. Moreover, it entails the need to take into account \textit{a priori} a sufficiently large number of spin-free electronic states 
for the evaluation of the spin-orbit Hamiltonian matrix elements 
in order to ensure convergence of the SO-coupled eigenstates. 
Hence, the predictive potential of such two-step approaches is inevitably limited for a finite set of spin-free states. 
By contrast, a genuine relativistic DMRG model, initially 
proposed for a ``traditional" (i.e. non-MPO based formulation) DMRG model in Ref.~\citenum{knec14}\ and proposed in this work within an efficient MPS/MPO framework of the DMRG algorithm, addresses simultaneously all of the above issues by providing access to 
large active orbital spaces \textit{combined} with a variational 
description of relativistic effects in the orbital basis. 

The paper is organized as follows: We first briefly review the concept of representing a wave function and an operator in a matrix-product \textit{ansatz}, 
respectively, including a short summary of the variational optimization of an MPS wave function in such a model. 
In Section \ref{sec:mps-rel-model}, a relativistic Hamiltonian framework suitable for a variational description of 
SO coupling is introduced and a detailed account of its implementation in an MPS/MPO setting of the 
DMRG \textit{ansatz}\ is given. In addition, we illustrate the calculation of first-order molecular properties within our relativistic DMRG model. 
Subsequently, after providing computational details in Section \ref{sec:mps-rel-comp-det}, we assess in Section \ref{sec:mps-rel-tlh} 
the numerical performance of our present implementation with respect to the calculation of absolute energies for the sample molecule thallium hydride. 
In Section \ref{sec:mps-rel-dy}, we demonstrate the applicability 
of our relativistic MPS/MPO-based DMRG model for the calculation of magnetic properties in a real-space approach 
at the example of a large Dy(III)-containing molecular complex. Conclusions and an outlook concerning future work based on the presented relativistic DMRG model are summarized 
in Section \ref{sec:conclusions}.

\section{Matrix product states and matrix product operators}\label{sec:mps-mpo-concepts}
We briefly introduce in this section the concepts of expressing a quantum state as an MPS and 
a (Hermitian) operator as an MPO in a non- or scalar-relativistic framework 
which will constitute the basis for our genuine relativistic DMRG model --- including a variational account of SO coupling --- discussed in Section \ref{sec:mps-rel-model}. 
Our notation closely follows the presentations of Refs.~\citenum{kell15a}\ and \citenum{kellerdiss}. 
A comprehensive review of the DMRG approach in an MPS/MPO formulation can be found in Ref.~\citenum{scho11}. 

\subsection{MPS}\label{sec:mps-mpo-concepts-mps}

\begin{figure}[tbph]
	\centering
	\includegraphics[width=0.55\textwidth]{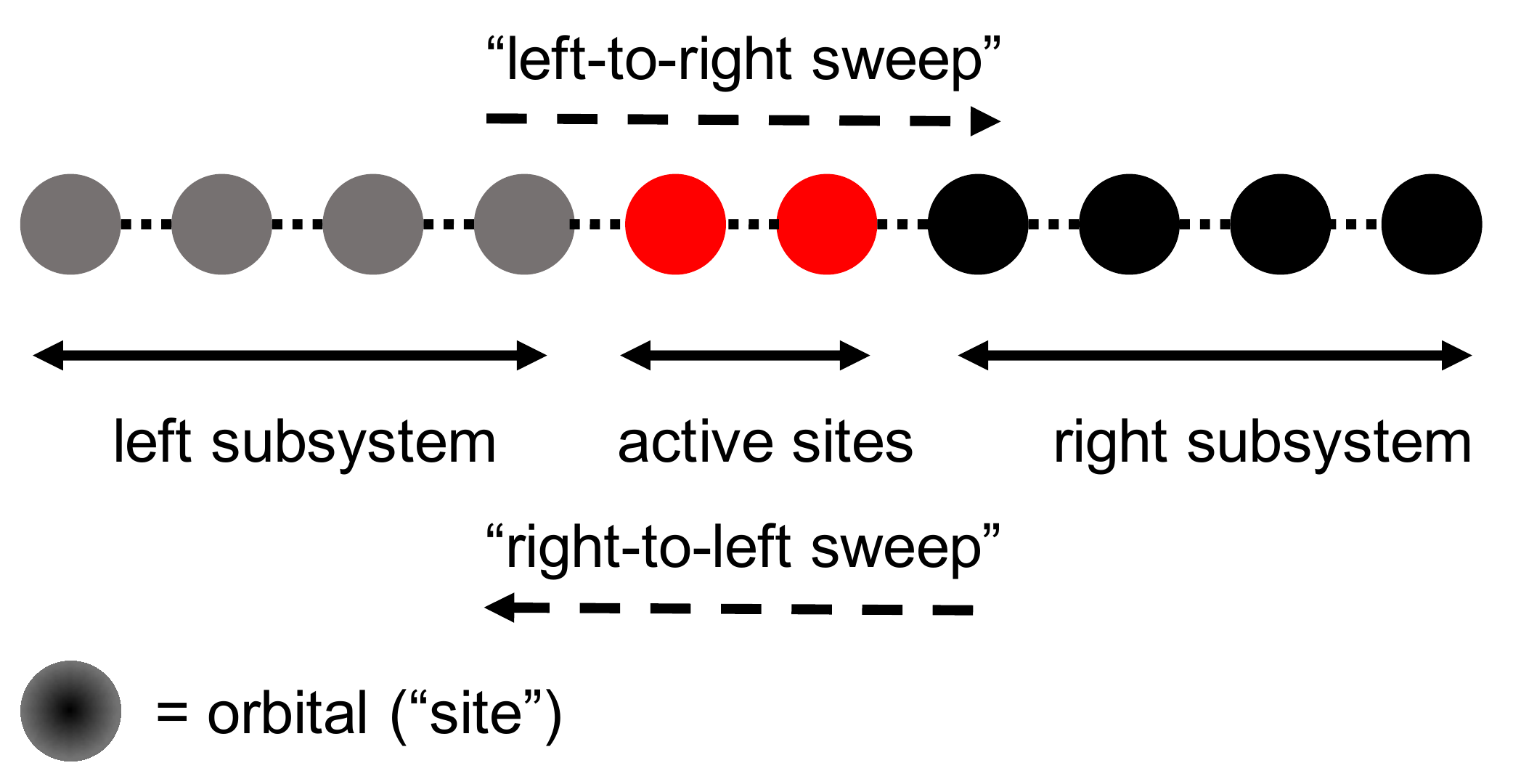}
	\caption{Alignment of orbitals (``sites") of a given active orbital space on a fictitious (as indicated by the connecting $\cdots$) one-dimensional lattice. The latter is partitioned into a left and right subsystem as well as active sites on which DMRG many-electron basis states are constructed and successively optimized in a variational approach. While traversing (``sweeping") through the lattice from left to right, the left subsystem is gradually enlarged by the left active site whereas the leftmost site on the right subsystem becomes active. Reaching the end of the lattice initiates a right-to-left sweep and the procedure is reversed.}
	\label{fig:mps-orbital-1D}
\end{figure}
Consider an arbitrary state $\kPsi$\ in a Hilbert space spanned by ($N$\ electrons distributed in) $L$\ spatial orbitals 
that are assumed to be ordered in a favorable fashion \cite{kell14} on a fictitious one-dimensional \textit{lattice}. 
In a traditional (CI-like) \textit{ansatz}, $\kPsi$ is commonly expressed as a linear superposition of occupation 
number vectors (ONVs) $\ket{\boldsymbol{k}}$\ with 
the expansion coefficients $c_{\ONstring}$ 
\begin{equation}\label{eq:CI_wave_function}
\kPsi = \sum\limits_{{\boldsymbol{k}}} c_{\boldsymbol{k}} \ket{\boldsymbol{k}} = \sum\limits_{\allStates} c_{\ONstring} \ONvec\ .
\end{equation}
Note that, in a non-relativistic formulation, 
each local space is of dimension four corresponding to the possible 
local basis states 
\begin{equation}\label{eq:local-space-dim}
k_l=\left\{\left|\uparrow\downarrow\right>,\left|\uparrow\right>,\left|\downarrow\right>,\left|0\right>\right\}\ ,
\end{equation}
for the $l$-th spatial orbital. 
By contrast, in an MPS representation of $\kPsi$, the CI 
coefficients $\{c_{\ONstring}\}$\ are written as a product of $m_{l-1}\times m_{l}$-dimensional matrices $M^{k_l} = \{M^{k_l}_{a_{l-1}a_l}\}$, i.e., four matrices corresponding to the local space dimension, 
\begin{align}
\kPsi &= \sum_{\allStates} \sum_{\alldim} M^{k_1}_{1 a_1} M^{k_2}_{a_1 a_2} \cdots M^{k_L}_{a_{L-1} 1} \ONvec = \sum_{\boldsymbol{k}} M^{k_1} M^{k_2} \cdots M^{k_L} \ket{\boldsymbol{k}}\ . \label{eq:MPS2}
\end{align} 
Given the $\{c_{\ONstring}\}$\ being scalar coefficients, requires that the final contraction of the matrices $M^{k_l}$\ must yield a scalar number. This further implies 
that the first and the last matrices in Eq.~(\ref{eq:MPS2}) are in practice $1\times m_1$-dimensional row and $m_{L-1}\times 1$-dimensional column vectors, respectively. 
The introduction of some maximum dimension $m$ for the matrices $M^{k_l}$, with $m$\ commonly 
referred to as \textit{number of renormalized block states} \cite{whit93}, constitutes the fundamental idea 
that facilitates to approximate a full CI-type wave function (cf.~Eq.~(\ref{eq:CI_wave_function}))  
to chemical accuracy while merely requiring to optimize a polynomial number of parameters in an 
MPS wave function \textit{ansatz}.

Moreover, the partitioning of the CI coefficients 
in products of the $M^{k_l}$\ matrices given in Eq.~(\ref{eq:MPS2})\ is not unique, i.e., $\kPsi$\ may be expressed (through successive applications of a singular value decomposition) in different ways such as in a left-canonical form \cite{scho11}
\begin{equation}
\label{eq:left-con}
	\kPsi =  \sum_{\boldsymbol{k}} A^{k_1} A^{k_2} \cdots A^{k_L} \ket{\boldsymbol{k}} ,
\end{equation}
with
\begin{equation}
\label{eq:left-norm-cond}
	\sum_{k_l} A^{k_l \dagger} A^{k_l} = I\ , 
\end{equation}
where $I$\ is the unit matrix and the $A^{k_l}$\ matrices are referred to as left-normalized MPS tensors. 
Similarly, the MPS can be written in right-canonical form \cite{scho11}
\begin{equation}
\label{eq:right-con}
	\kPsi = \sum_{\boldsymbol{k}} B^{k_1} B^{k_2} \cdots B^{k_L} \ket{\boldsymbol{k}}\ ,
\end{equation}
where the $B^{k_l}$ matrices are right-normalized, satisfying the condition
\begin{equation}
\label{eq:right-norm-cond}
	\sum_{\sigma_k} B^{k_l} B^{k_l \dagger} = I\ .
\end{equation}
The non-uniqueness, which allows us to express an MPS in different forms originates from the existence of a 
gauge degree of freedom \cite{scho11}. Considering for a given MPS two adjacent sets of 
matrices $M^{k_l}$\ and $M^{k_{l+1}}$\ with matching bond dimensions $m$, it can be shown \cite{scho11}\ that 
the MPS is invariant with respect to a right (left) multiplication with matrix $X$ of dimension $m \times m$,
\beq
 M^{k_l} \rightarrow  M^{k_l} X\,\,\, \forall k_l, \qquad\  M^{k_{l+1}} \rightarrow  X^{-1} M^{k_{l+1}}\,\,\, \forall k_l\ ,
\label{eq:mps-gauge-freedom}
\eeq
provided that $X$\ is invertible and non-singular.

The gauge freedom may therefore be further exploited to express the MPS of Eq.~(\ref{eq:MPS2}) 
in mixed-canonical form at sites (orbitals) $\{l, l+1\}$ \cite{scho11}
\begin{equation}\label{PsiMPScan}
 \ket{\Psi} = \sum_{\boldsymbol{k}} \sum_{a_1,\cdots,a_{L-1}}A_{1 a_1}^{k_1}\cdots A_{a_{l-2} a_{l-1}}^{k_{l-1}} M_{a_{l-1},a_{l+1}}^{k_l k_{l+1}} B_{a_{l+1} a_{l+2}}^{k_{l+2}}\cdots B_{a_{L-1} 1}^{k_L} \ket{\boldsymbol{k}}\ ,
\end{equation}
where the MPS tensors $A^{k_{l-1}}=\{A^{k_{l-1}}_{a_{l-2}a_{l-1}}\}$\ and $B^{k_{l-1}}=\{B^{k_{l+2}}_{a_{l+1}a_{l+2}}\}$\ are left- and right-normalized, respectively, as shown in Eqs.~(\ref{eq:left-norm-cond})\ and (\ref{eq:right-norm-cond})\ and $M_{a_{l-1},a_{l+1}}^{k_l k_{l+1}}$\ is a two-site MPS tensor
\begin{equation}\label{MPStwosite}
M_{a_{l-1},a_{l+1}}^{k_l k_{l+1}} = \sum\limits_{a_l} M_{a_{l-1},a_{l}}^{k_l} M_{a_{l},a_{l+1}}^{k_{l+1}}\ .
\end{equation}
The mixed-canonical form will be a central element of the two-site DMRG optimization algorithm outlined in Section \ref{sec:mps-mpo-concepts-varopt}. 

\subsection{MPO}\label{sec:mps-mpo-concepts-mpo}

Exploiting the matrix-product formulation introduced in the previous section for operators, an $N$-electron 
operator $\widehat{W}$\ can be expressed in MPO form as
\begin{eqnarray}\label{eq:MPO}
\widehat{\mathcal{W}} & = & \sum_{\allStates} \sum_{\allStatesopp} \sum_{\alldimop} W^{k_1 k_1^{\prime}}_{1 b_1} W^{k_2 k_2^{\prime}}_{b_1 b_2} \cdots W^{k_L k_L^{\prime}}_{b_{L-1} 1} 
\ONvec \ONvecb\nonumber\\
&\equiv& \sum_{\boldsymbol{k} \boldsymbol{k^{\prime}}} w_{\boldsymbol{k} \boldsymbol{k^{\prime}}} \ket{\boldsymbol{k}}\bra{\boldsymbol{k^{\prime}}}\ ,
\end{eqnarray}
with the incoming and outgoing physical states $k_l$\ and $k_l^{\prime}$\ and the virtual indices $b_{l-1}$\ and $b_l$. 
Rearranging the summations and changing the order of contraction in Eq.~({\ref{eq:MPO}}) leads to \cite{kell15a,kell16}
\begin{equation}\label{eq:mpoMOD}
\widehat{\mathcal{W}}  = \sum\limits_{\alldimop} W^{1}_{1 b_{1}} \cdots W^{l}_{b_{l-1} b_{l}} \cdots W^{L}_{b_{L-1} 1}\ ,
\end{equation}
where the entries of the $\left\{{W}^{l}_{b_{l-1} b_{l}}\right\}$\ matrices 
comprise only \textit{local} creation (annihilation) operators acting on the $l$-th orbital. 
For example, consider the operator $\tilde{a}_{\uparrow_l}^{\dagger}$\ which can be written 
as a linear combination of the local basis states 
\begin{equation}\label{eq:create-sum}
\tilde{a}_{\uparrow_l}^{\dagger} = \left|\uparrow \downarrow\right>\left<\downarrow\right| + \left|\uparrow\right>\left<0\right|\ .
\end{equation}
Hence, Eq.~(\ref{eq:create-sum}) implies that the matrix representation of $\tilde{a}_{\uparrow_l}^{\dagger}$\ 
corresponds to a $(4\times4)$-dimensional matrix with two non-zero entries equal to one. 
Likewise, similar considerations hold for the remaining local operators while 
the matrix representations of the corresponding \textit{local, relativistic} operators of our relativistic DMRG model 
will be discussed in Section \ref{sec:mps-rel-model-mpo}.

For completeness, we conclude this section by considering the action of the local Hamiltonian $\hat{H}$ given in MPO form on an MPS state in mixed-canonical form (cf.~Eq.~(\ref{PsiMPScan})) at sites \{$l,l+1$\}\ which can be written as \cite{scho11} 
\begin{equation}\label{HPsiMPS}
\hat{H} \kPsi = \sum_{b_{l-1},b_l} \sum_{a'_{l-1},k'_l,a'_l} L_{b_{l-1}}^{a_{l-1},a'_{l-1}} W_{b_{l-1},b_{l+1}}^{k_l k_{l+1},k'_l k'_{l+1}} R_{b_{l+1}}^{a_{l+1},a'_{l+1}} M_{a_{l-1},a_{l+1}}^{k'_l,k'_{l+1}} | a_{l-1} \rangle_A |k_l \rangle |k_{l+1} \rangle |a_{l+1} \rangle_B\ ,
\end{equation}
with the left and right basis states given by
\begin{equation}\label{PsiMPSA}
 | a_{l-1} \rangle_A = \sum_{k_{1},\cdots,k_{l-1}} \sum_{a_1,\cdots,a_{l-1}}A_{1 a_1}^{k_1}\cdots A_{a_{l-2} a_{l-1}}^{k_{l-1}} |k_1,\cdots,k_{l-1} \rangle\ ,
\end{equation}
and 
\begin{equation}\label{PsiMPSB}
 | a_{l+1} \rangle_B = \sum_{k_{l+1},\cdots,k_{L}} \sum_{a_{l+1},\cdots,a_{L}} B_{a_{l+1} a_{l+2}}^{k_{l+2}}\cdots B_{a_{L-1} 1}^{k_L} |k_{l+1},\cdots,k_{L} \rangle\ .
\end{equation} 
Here, we introduced the left and right boundaries as \cite{scho11,kell15a}
\begin{equation}\label{Lcontr}
 L_{b_{l-1}}^{a_{l-1},a'_{l-1}} = \sum_{\{a_i,b_i,a'_i;i<l-1\}} \left( \sum_{k_1,k'_1} A_{1,a_1}^{k_1 \ast} W_{1,b_1}^{k_1,k'_1}A_{1,a'_1}^{k'_1} \right )\cdots \left( \sum_{k_{l-1},k'_{l-1}} A_{a_{l-2},a_{l-1}}^{k_{l-1} \ast} W_{b_{l-2},b_{l-1}}^{k_{l-1},k'_{l-1}} A_{a'_{l-2},a'_{l-1}}^{k'_{l-1}} \right)\ ,
\end{equation}
and 
\begin{eqnarray}\label{Rcontr}
 R_{b_{l+1}}^{a_{l+1},a'_{l+1}} &=& \sum_{\{a_i,b_i,a'_i;i>l+1\}} \left( \sum_{k_{l+1},k'_{l+1}} B_{a_{l+1},a_{l+2}}^{k_{l+2}\ast} W_{b_{l+1},b_{l+2}}^{k_{l+2},k'_{l+2}} B_{a'_{l+1},a'_{l+2}}^{k'_{l+2}} \right ) \cdots \nonumber \\
 & & \times \left( \sum_{k_{L},k'_{L}} B_{a_{L-1},a_1}^{k_{L} \ast} W_{b_{L-1},b_{1}}^{k_{L},k'_{L}} B_{a'_{L-1},a'_{1}}^{k'_{L}} \right )\ ,
\end{eqnarray}
as well as the two-site MPO tensor \cite{scho11,kell15a}
\begin{equation}\label{MPOtwosite}
W_{b_{l-1},b_{l+1}}^{k_l k_{l+1} k'_l k'_{l+1}} = \sum\limits_{b_l} W_{b_{l-1},b_{l}}^{k_l k'_l} W_{b_{l},b_{l+1}}^{k_{l+1} k'_{l+1}}\ .
\end{equation}

\subsection{Variational optimization of an MPS}\label{sec:mps-mpo-concepts-varopt}

In a variational optimization of $\ket{\Psi}$\ with the Hamiltonian $\hat{H}$\ expressed in MPO form, the objective is to 
minimize the energy expectation value $\left<\Psi | \hat{H}  |\Psi\right>$ with respect to the entries of the MPS tensors under 
the constraint that the wave function is normalized, i.e., $\left<\Psi|\Psi\right> = 1$. To this end, 
a Lagrangian multiplier $\lambda$ is introduced to ensure normalization. The optimization of the MPS then corresponds to find the extremum of the Lagrangian \cite{chan08b}
\begin{equation} \label{eq:gs_func}
	\mathcal{L} = \bra{\Psi}\hat{H}\ket{\Psi} - \lambda\left<\Psi\right|\left.\Psi\right> .
\end{equation}
The optimization parameters are the matrices $M^{k_l}$ of the MPS. To solve this highly non-linear optimization problem, 
the same idea as in the original (non-MPO) DMRG algorithm is adopted: iterating through the lattice and optimizing the entries $M_{a_{l-1,a_l}}^{k_l}$ of one matrix (single-site variant) or of the two-site MPS tensor $M_{a_{l-1},a_{l+1}}^{k_l,k_{l+1}}$ (two-site variant) at a time, while keeping all the others fixed. This approach not only greatly simplifies the optimization problem to one of lower complexity but also ensures to find a better approximation to the true ground state in a variational sense. 

In the following, we consider the two-site variant (corresponding to treating simultaneously two sites as ``active" 
as indicated by a red color code in Figure \ref{fig:mps-orbital-1D}). Taking the derivative with respect to the complex conjugate of $M_{a_{l-1},a_{l+1}}^{k_l,k_{l+1}}$\ 
\begin{equation}
	\frac{\partial}{\partial M_{a_{l-1},a_{l+1}}^{k_l,k_{l+1}\ast}} (\bra{\Psi}\hat{H}\ket{\Psi} - \lambda\left<\Psi\right|\left.\Psi\right>) = 0\ ,
\end{equation}
yields \cite{scho11}
\begin{equation} \label{eq:ts_optimization}
\begin{split}
\sum_{a'_{l-1} a'_{l},b_{l-1} b_{l+1}} \sum_{k'_l k'_{l+1}} L_{b_{l-1}}^{a_{l-1},a'_{l-1}} W_{b_{l-1},b_{l+1}}^{k_l k_{l+1},k'_l k'_{l+1}} R_{b_{l+1}}^{a'_{l+1},a_{l+1}} M_{a'_{l-1},a'_{l+1}}^{k'_l k'_{l+1}} = \lambda M_{a_{l-1},a_{l+1}}^{k_l k_{l+1}}\ ,
\end{split}
\end{equation}
where we assumed that the left and right boundaries (cf.~Eqs.~(\ref{Lcontr})\ and (\ref{Rcontr})) were calculated from 
left- and right-normalized MPS tensors, respectively.

The resulting Eq.~(\ref{eq:ts_optimization}) can be recast in a matrix eigenvalue equation \cite{ostl95,scho11}\ 
\begin{equation}\label{EigPsiMPS}
\boldsymbol{\mathcal{{H}}} v - \lambda v = 0\ ,
\end{equation}
with the local Hamiltonian matrix $\boldsymbol{\mathcal{{H}}}$ at sites \{$l,l+1$\}\ after reshaping given by 
\begin{equation}\label{EigPsiMPSH}
H_{(k_{l,l+1} a_{l-1}a_{l+1}),(k'_{l,l+1} a'_{l-1}a'_{l+1})}  =  \sum_{b_{l-1},b_{l+1}} L_{b_{l-1}}^{a_{l-1},a'_{l-1}} W_{b_{l-1},b_{l+1}}^{k_l k_{l+1},k'_l k'_{l+1}} R_{b_{l+1}}^{a'_{l+1},a_{l+1}}\ ,
\end{equation}
and the vector $v$ collecting
\begin{equation}\label{EigPsiMPSV}
v_{k_l k_{l+1} a_{l-1} a_{l+1}} = M_{a_{l-1},a_{l+1}}^{k_l k_{l+1}}\ .
\end{equation}

Since we are often interested in only a few of the lowest eigenvalues $\lambda$, Eq.~(\ref{EigPsiMPS}) is best solved by an iterative eigensolver such as 
the Jacobi-Davidson procedure. For example, having obtained the lowest eigenvalue $\lambda_0$ and the corresponding eigenvector 
$v^0_{k_l k_{l+1} a_{l-1} a_{l+1}}$, the latter can be reshaped back to $M_{a_{l-1},a_{l+1}}^{k_l k_{l+1}}$\ which is then subject to a left- or right-normalization into $A^{k_l}_{a_{l-1}a_{l}}$ or $B^{k'_{l+1}}_{a_{l}a_{l+1}}$ by a singular value decomposition (discarding the 3$m$ smallest singular values) 
in order to maintain the desired normalization structure and the dimensionality of the MPS tensors. Given the optimized MPS tensors for sites $l$\ and $l+1$, the complete algorithm now \textit{sweeps}\ sequentially forth and back 
from left-to-right and right-to-left through the {lattice} (cf.~Figure \ref{fig:mps-orbital-1D}) consisting of the $L$\ spatial orbitals ordered in a suitable form. At each step the MPS tensors  $\{M^{k_l}_{a_{l-1}a_l}\}$\ are optimized until convergence is reached.

\section{Formulation of a relativistic DMRG model}\label{sec:mps-rel-model}

\subsection{Hamiltonian framework}\label{sec:rel-ham}

In accordance with the pilot relativistic DMRG implementation presented in Ref.~\citenum{knec14}, our present formulation of 
a relativistic DMRG model in a matrix-product \textit{ansatz} is based on the electronic, time-independent, external magnetic-field-free   
Dirac-Coulomb(-Breit) Hamiltonian (note that any two-component Hamiltonian is equally well applicable) which encompasses all 
important relativistic ``corrections" \cite{dyal07,reih09,auts12}\ relevant for chemistry.  
Invoking the no-pair approximation, this operator can be written in second quantization as \cite{jens96,dyal07}, 
\beqa 
\hat{H}_{} & = & \sum_{pq}\left[h_{pq}{\hat{X}_{pq}^{+}} +
\frac{1}{2} \left(h_{\bar{p}q}{\hat{X}_{\bar{p}q}^{+}} +
h_{p\bar{q}}{\hat{X}_{p\bar{q}}^{+}} \right)\right] \nonumber \\ & & +
\frac{1}{2} \sum_{pqrs}\left[\left(pq|rs\right)\hat{x}^{++}_{pq,rs} +
\left(\bar{p}q|rs\right)\hat{x}^{++}_{\bar{p}q,rs} +
\left(p\bar{q}|rs\right)\hat{x}^{++}_{p\bar{q},rs} \right] \nonumber \\
&
& +
\frac{1}{4} \sum_{pqrs}
\left(\bar{p}q|r\bar{s}\right)\hat{x}^{++}_{\bar{p}q,r\bar{s}} \nonumber
\\ & &
+ \frac{1}{8} \sum_{pqrs}
\left[\left(\bar{p}q|\bar{r}s\right)\hat{x}^{++}_{\bar{p}q,\bar{r}s} +
\left(p\bar{q}|r\bar{s}\right)\hat{x}^{++}_{p\bar{q},r\bar{s}} \right] \nonumber \\
&&
-\frac{1}{2} \sum_{pqrs}\left[\left(p\boldsymbol{\alpha}q|r\boldsymbol{\alpha}s\right)\hat{x}^{--}_{pq,rs} +
 \left(\bar{p}\boldsymbol{\alpha}{q}|r\boldsymbol{\alpha}s\right)\hat{x}^{--}_{\bar{p}q,rs} + 
 \left({p}\boldsymbol{\alpha}\bar{q}|r\boldsymbol{\alpha}s\right)\hat{x}^{--}_{p\bar{q},rs} \right] \nonumber \\
 &&
 -\frac{1}{4} \sum_{pqrs}
 \left(\bar{p}\boldsymbol{\alpha}q|r\boldsymbol{\alpha}\bar{s}\right)\hat{x}^{--}_{\bar{p}q,r\bar{s}} \nonumber
 \\ & &
 - \frac{1}{8} \sum_{pqrs}
 \left[\left(\bar{p}\boldsymbol{\alpha}q|\bar{r}\boldsymbol{\alpha}s\right)\hat{x}^{--}_{\bar{p}q,\bar{r}s} +
 \left(p\boldsymbol{\alpha}\bar{q}|r\boldsymbol{\alpha}\bar{s}\right)\hat{x}^{--}_{p\bar{q},r\bar{s}} \right]\ ,
\label{hamilrel}
\eeqa
where the last three rows on the right-hand side of Eq.~\eqref{hamilrel} comprise the two-electron terms originating from the Breit (Gaunt) 
interaction and the summation indices $p,q,r,s$\ strictly refer to positive-energy spinors. 
Moreover, following earlier works \cite{jens96,thys08,kim13,knec14,bate15}, 
we assume in Eq.~\eqref{hamilrel}\ a Kramers pair basis $\{\varphi_l,\bar{\varphi}_l\}$\ built from  
pairs of fermion spinor functions $\{\varphi_l\}$\ and $\{\bar{\varphi}_l\}$, respectively. This implies that the summations in Eq.~\eqref{hamilrel} run 
over Kramers pairs and not single spinors. According to Kramers theorem, which holds in the absence of an external magnetic field, 
states of a single fermion function $\{\varphi_l\}$\ are (at least) doubly degenerate where each state is related to its energy-degenerate 
partner by time reversal \cite{dyal07,reih09}, i.e., 
\beq
\hat{\mathcal{K}}\varphi_l = \bar{\varphi}_l\ ,
\eeq
where $\hat{\mathcal{K}}$ is the time-reversal operator. 
In addition, we employ in Eq.~\eqref{hamilrel}\ 
Kramers single- ${\hat{X}}^s_{pq}$\ (with $s = \pm$) and double replacement 
operators ${\hat{x}}^{s_1,s_2}_{pq,rs}$\ (with $s_1,s_2 = \pm$) \cite{aucar}, respectively. They can be expressed 
in terms of creation and annihilation operators as \cite{aucar,jens96,flei01,dyal07}
\beq
{\hat{X}}^s_{pq} = a_p^{\dag} a_q + sa_{\bar{q}}^{\dag} a_{\bar{p}}
\qquad {\hat{X}}^s_{{\bar{p}}q} = a_{\bar{p}}^{\dag} a_{{q}} - sa_{\bar{q}}^{\dag}
a_{{p}} \qquad 
{\hat{X}}^s_{p{\bar{q}}} = a_{{p}}^{\dag} a_{\bar{q}} - sa_{{q}}^{\dag} a_{\bar{p}}\ ,
\label{eq:kramx-one}
\eeq
and 
\beq
{\hat{x}}^{s_1,s_2}_{pq,rs} = {\hat{X}}^{s_1}_{pq} {\hat{X}}^{s_2}_{rs} - \delta_{rq} a_p^{\dag}a_s -
s_1 \delta_{r{\bar{p}}} a_{\bar{q}}^{\dag} a_s -
s_2 \delta_{{\bar{s}}q} a_{{p}}^{\dag} a_{\bar{r}} - s_1 s_2
\delta_{{\bar{p}}{\bar{s}}} a_{\bar{q}}^{\dag} a_{\bar{r}} = {\hat{x}}^{s_2,s_1}_{rs,pq}\ ,
\label{eq:kramx-two}
\eeq
where the sign indices $s$\ and ${{s_{1}},{s_{2}}}$ in Eqs.~\eqref{eq:kramx-one} and \eqref{eq:kramx-two} 
indicate the symmetry of the operators under time 
reversal and Hermitian conjugation, respectively. The remaining double replacement operators appearing in Eq.~\eqref{hamilrel}\ 
follow from Eq.~\eqref{eq:kramx-two}] through the application 
of time-reversal symmetry to the creation (annihilation) operators \cite{dyal07}.

An efficient account of double-group symmetry is a crucial aspect of 
our relativistic DMRG formulation. In contrast to a spin-free approach, where spin and spatial degrees of freedom can be considered independently, 
SO coupling introduced through the relativistic molecular 
Hamiltonian (cf.~Eq.\eqref{hamilrel}), entails a coupling of both degrees of freedom. 
Hence, rather than having to deal with simple spatial point group symmetry we have to consider the corresponding double groups. In contrast to the single 
groups, these groups comprise extra irreps originating from an 
introduction of a rotation 2$\pi$ about an arbitrary axis \cite{tink64,dyal07,reih09}. 
These additional irreps are commonly referred to as fermion irreps and 
are spanned by spinors whereas the ``regular" ones are called boson 
irreps spanned, for example, by spinor products and operators.  

Following the symmetry handling of our pilot relativistic 
DMRG implementation \cite{knec14}, we also take advantage of a quaternion 
symmetry scheme introduced by Saue and Jensen \cite{saue99}\ in the \dir\ program package \cite{DIRAC12}\ to which our DMRG software \qcm\ is interfaced. In addition to the Abelian subgroups of the binary double D$_{2h}^\ast$, 
we introduced in this work the double groups C$_{16h}^\ast$\ and 
C$_{32v}^{\ast}$\ for linear molecules with and without inversion symmetry as finite Abelian approximations to the full groups D$_{\infty h}^{\ast}$\ and C$_{\infty v}^{\ast}$, respectively.  
Exploiting a quaternion symmetry scheme allows us to resort to 
either real (for the double groups D$_{2h}^{\ast}$, D$_{2}^{\ast}$\ and C$_{2v}^{\ast}$), complex (C$_{2h}^{\ast}$, C$_{2}^{\ast}$\ and C$_{s}^{\ast}$)\ or quaternion algebra (C$_{i}^{\ast}$\ and C$_{1}^{\ast}$)\ in the construction of a matrix representation 
of the Hamiltonian operator (cf.~Eq.\eqref{hamilrel})\ in a finite 
Kramers-paired spinor basis and the subsequent solution of the 
corresponding eigenvalue problem \cite{saue99,dyal07}, respectively.  
Moreover, working in a Kramers-paired spinor basis, 
entails that, by symmetry, all matrix elements of a 
time-reversal symmetric one-electron operator 
with an odd number of barred spinor indices are zero 
\cite{saue99}. Furthermore, for the two-electron integrals, we exploit the so-called (\texttt{NZ},3) matrix representation \cite{thyssendiss,thys08}\ where the number of nonzero rows for a given binary double group  corresponds to the rank \texttt{NZ}\ of the matrix.  
For example, as shown in Eq.~\eqref{hamilrel}, 
a two-electron integral $(pq|rs)$\ may generally
comprise a mixed number of barred and unbarred spinors. 
Following the (\texttt{NZ},3) classification discussed in detail in Ref.~\citenum{thys08}, 
for real- and complex-valued double groups, labeled as \texttt{NZ=1}\ and \texttt{NZ=2}, respectively, only integrals with an even number ($n_{\rm barred}=0,2,4$) 
of barred spinors are non-vanishing. By contrast, for quaternion-valued double 
groups with \texttt{NZ}=4, all integrals are in general non-zero.

\subsection{Relativistic DMRG within an MPS/MPO framework}\label{sec:mps-rel-model-mpo}

In the previous section \ref{sec:rel-ham}, we briefly introduced a Hamiltonian framework suitable for a relativistic DMRG model. 
In the following we will elaborate on details of a relativistic DMRG model which exploits an MPS wave function and MPO Hamiltonian representation. The basic concepts are identical to those discussed in Section \ref{sec:mps-mpo-concepts}\ for a non- or scalar-relativistic framework. Hence, we will in particular focus on differences that originate from the underlying use of a relativistic Hamiltonian model. We first provide a brief, general introduction to the implementation of (double group) symmetry in our MPS/MPO framework. Subsequently, 
details concerning the actual implementation of a relativistic DMRG model in our \qcm\ software packages are presented. 

\subsubsection{Symmetry} \label{sec:mps-rel-model-mpo-symmetry}

As indicated in Section \ref{sec:mps-rel-model}, 
exploiting symmetries in a (relativistic) DMRG algorithm is a key element to 
enhance both accuracy and computational efficiency (see for example also Ref.~\citenum{kell16}\ for the spin-symmetry adaptation of the MPS/MPO representation of wave function and operators in our \qcm\ software package). 
To illustrate the account of symmetry in our relativistic DMRG model, we consider an operator $\hat{Q}$\ that commutes with the Hamiltonian of Eq.~(\ref{hamilrel}), i.e.,
\begin{equation}\label{eq:qhcomm}
\left[ \hat{H} , \hat{Q} \right] = 0\ .
\end{equation}
Hence, the eigenfunctions $\kPsi$\ of the Hamiltonian can always be chosen as eigenfunctions of $\hat{Q}$\ with 
\begin{equation}\label{eq:qeig}
\hat{Q} \kPsi = q \kPsi\ .
\end{equation}

Hence, it is a \textit{sine qua non} in the DMRG optimization algorithm that in each step any local operation 
transforms the basis according to the group it belongs to such that any element of the basis 
remains an element of the group after the transformation. 
To further illustrate this concept, we consider two operators $\hat{Q}_1$ and $\hat{Q}_2$\ which both commute with $\hat{H}$, 
e.g., 
\begin{align}
\left[ \hat{H} , \hat{Q}_1 \right] &= 0 , \\
\left[ \hat{H} , \hat{Q}_2 \right] &= 0 ,\ 
\end{align}
with the corresponding eigenvalues (in the following denoted as \textit{quantum numbers}) 
given by 
\begin{align}
\hat{Q}_1 \kPsi &= \texttt{q}_1 \kPsi , \label{eq:q1}\\
\hat{Q}_2 \kPsi &= \texttt{q}_2 \kPsi . \label{eq:q2}
\end{align}
This allows us to label the eigenstate $\kPsi$ according to the quantum numbers given by Eqs.~(\ref{eq:q1}) and (\ref{eq:q2}), respectively, 
\begin{equation}\label{eq:psilabel}
\kPsi \rightarrow \texttt{<q$_1$,q$_2$>:j}\  ,
\end{equation}
where $\texttt{j}$ is an integer counting the possible number of realizations for this state with quantum numbers \texttt{q$_1$} and \texttt{q$_2$}\ during the DMRG optimization procedure. 
Assume, for example, in a DMRG sweep starting with the left subsystem composed by only the first site and with the symmetries of the total system defined by the particle number \texttt{N} and a given double group symmetry. 
Hence, $\hat{Q}_1 = \hat{N}$ and $\hat{Q}_2 = \hat{G}$, where $\hat{G}$ is any allowed operation of the double group. The quantum numbers are therefore relabeled as $\texttt{q}_1 \rightarrow \texttt{N}$ and $\texttt{q}_2 \rightarrow \texttt{g}$. 
At each site, the local space has dimension two, with one state corresponding to an empty spinor (labeled \texttt{<1,0>:1}) and the other state corresponding to an occupied spinor (labeled \texttt{<1,g}$_l$\texttt{>:1}), where \texttt{g}$_l$ refers to the irreducible representation of the spinor $\varphi_l$\ on site $l$. 
Note that the lattice is entered from the left with the vacuum state $\ket{a_0} = \texttt{<0,0>:1}$. After optimization of the first two sites, the first site 
is merged into the left subsystem block consisting of the vacuum state such that the new basis states are given by
\begin{equation}
\lbrace \ket{a_1} \rbrace = \ket{a_0} \otimes \lbrace \ksig[1] \rbrace .
\end{equation}
The two states defined on the first site are both eigenstates of the symmetry operators and can be labeled accordingly as
\begin{align}
\ket{k_1 = 0} &= \texttt{<0,0>:1}\ , \\
\ket{k_1 = 1} &= \texttt{<1,g$_1$>:1}\ .
\end{align}
Since the state $\ket{a_0}$ is an eigenstate of the symmetry operators, the tensor product will also be an eigenstate. This condition can always be enforced by the following equalities
\begin{align}
\hat{N} \ket{a_0} \ast \hat{N} \ksig[1] &= \hat{N} \ket{a_1}\ ,\\
\hat{G} \ket{a_0} \ast \hat{G} \ksig[1] &= \hat{G} \ket{a_1}\ ,
\end{align}
where $\ast$ denotes the group operation. Including the second site into the left subsystem 
results again in a basis of eigenstates of the symmetry operations, 
\begin{align}
\hat{N} \ket{a_1} \ast \hat{N} \ksig[2] &= \hat{N} \ket{a_2}\ ,\\
\hat{G} \ket{a_1} \ast \hat{G} \ksig[2] &= \hat{G} \ket{a_2}\ ,
\end{align}
where, by construction, the states $\{\ket{a_1}\}$ are eigenstates of the symmetry operators. 
The new states of the enlarged left block comprising the first two sites are therefore eigenstates of $\hat{N}$ and $\hat{G}$. 
This procedure can be iterated until the end of the lattice is reached which will lead to a basis set of many-particle states that satisfy 
the global symmetry constraints.

The above considerations hold for any type of parametrization of the quantum states, hence, also for an MPS wave function representation. 
To illustrate the latter, we consider a lattice of $L$ spinors and impose as symmetry constraints a total number of particles of $\texttt{N}=4$ and the totally symmetric irreducible representation for the target quantum state, i.e., \texttt{g}=0. 
Following the procedure outlined above for all sites $4< l \leq L$\ all states arising from the tensor product of the left subsystem 
with the local $l$-th site basis that would lead at the final site $L$\ to $\texttt{N}>4$ will be automatically discarded 
because they do not satisfy our imposed symmetry constraints for the target state. The same restrictions hold for all intermediate 
states that satisfy $\texttt{N}=4$\ but, combined in all possible ways with the corresponding local states up to site $L$, would yield $\texttt{g}\neq0$\ because the resulting quantum state would no longer transform as the totally symmetric irreducible representation. 
Hence, exploiting symmetries leads to a considerable reduction of the number of possible states that need to be taken into account which in turn lowers the associated numerical effort of an MPS wave function optimization.

\subsubsection{Implementation aspects}\label{sec:implementation}

\begin{figure}[tbh]
\centering
		\includegraphics[scale=0.60]{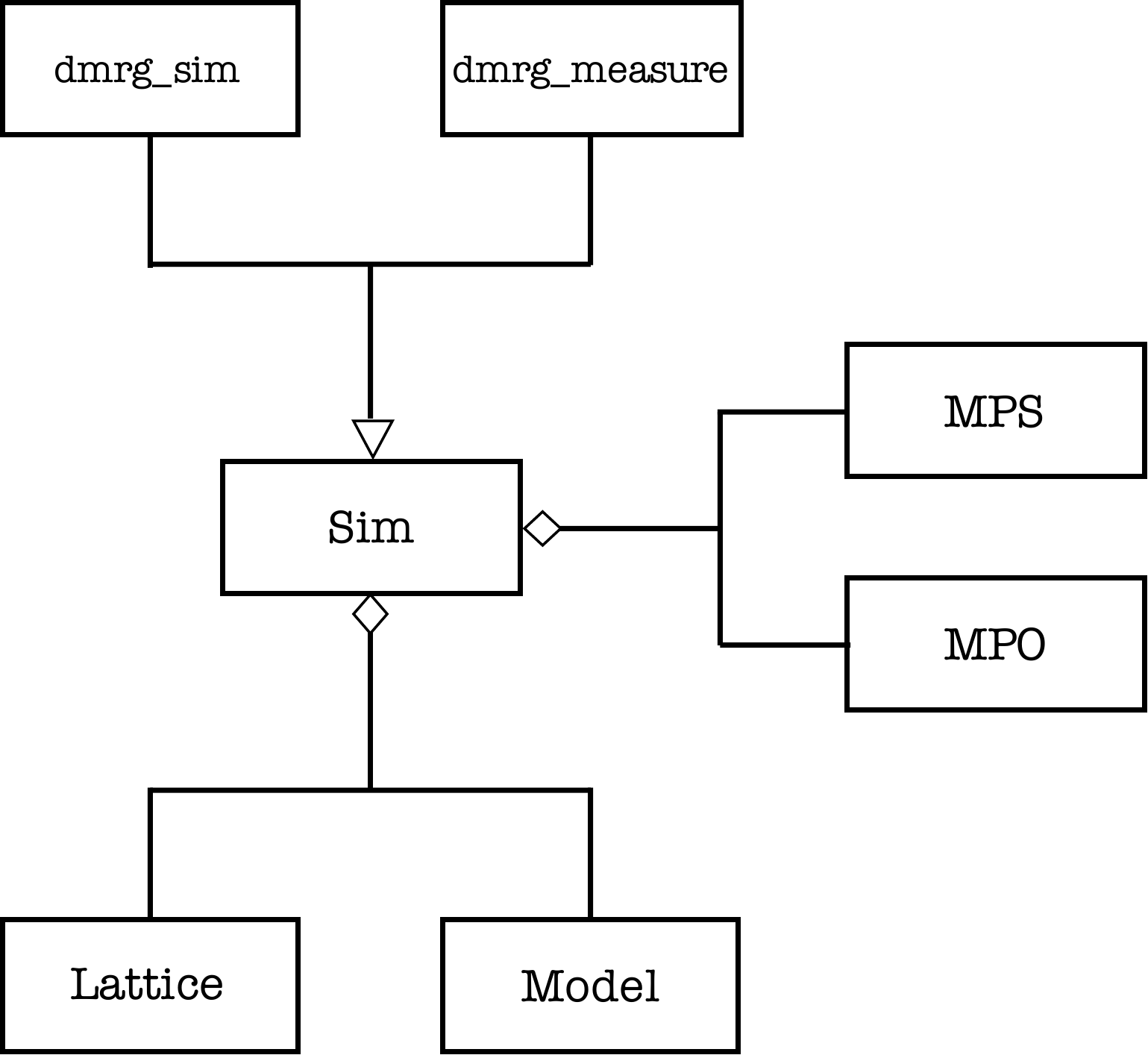}
	\caption{UML class diagram outlining the logic of the \qcm\ program.}
	\label{fig:class_diagram_logic}
\end{figure}

In this section we discuss selected important aspects of the actual implementation of 
our relativistic Hamiltonian model within the existing (nonrelativistic) 
framework of \qcm, while further details on the building blocks of \qcm\ can be found in Refs.~\citenum{dolf14,kell15a,kell16,kellerdiss}. 
Figure \ref{fig:class_diagram_logic}\ illustrates the main classes that constitute the essential building blocks for an 
MPS wave function optimization based on a given Hamiltonian model within the framework of \qcm. 

\begin{figure}[tbh]
\centering
	 \includegraphics[scale=0.3]{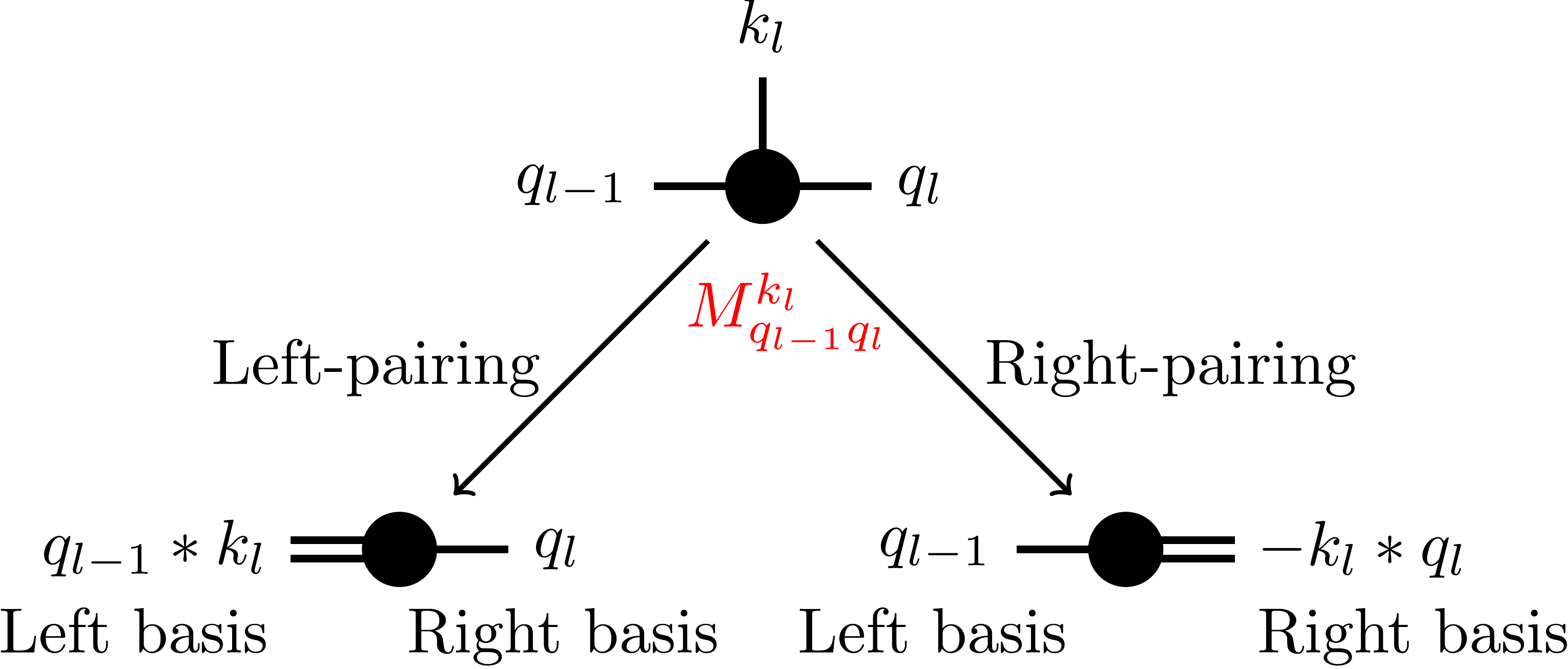}
	\caption{Left- and right-paired MPS tensors. Left-pairing implies the physical index to be fused with the left index whereas 
		right-pairing indicates a fusion of the inverse of the physical index with the right index.}
	\label{fig:left_right_paired}
\end{figure}

The central objects in \qcm\ are the MPS tensors $\left\{M^{k_l}\right\}$, introduced in Section \ref{sec:mps-mpo-concepts-mps}, 
which are represented by the class \texttt{MPS}\ in Figure \ref{fig:class_diagram_logic}. By taking into account the symmetry considerations outlined in the previous section, 
we can associate each MPS tensor $a$-index (cf.~Eq.~\eqref{eq:MPS2}) with a quantum number (or \texttt{charge}) \cite{kellerdiss}
\begin{equation}\label{eq:quant-num-for-MPS}
q_l = <\texttt{N}_l,\texttt{g}_l>\ ,
\end{equation}
such that the MPS tensor $M^{k_l}_{q_{l-1} a_{l-1};q_l a_l}$ is characterized by the symmetry constraint
\begin{equation}\label{eq:quant-num-for-MPS-constr}
q_l \in q_{l-1} \otimes k_l\ . 
\end{equation}
Hence, an MPS tensor is labeled by the quantum numbers $q_{l-1}$\ and $q_l$\ and $k_l$\ which are also referred 
to as \textit{left}, \textit{right}, and \textit{physical index}, respectively \cite{kellerdiss}. 
The resulting MPS tensor therefore exhibits a block structure with one block matrix for each of the local basis states of $k_l$. 
For an efficient storage, the physical index is \textit{fused} in \qcm\ 
either to the left or right index resulting in a rank-two tensor, i.e., a matrix, as depicted in Figure \ref{fig:left_right_paired}.

\begin{figure}[tbh]
\centering
	 \includegraphics[scale=0.35]{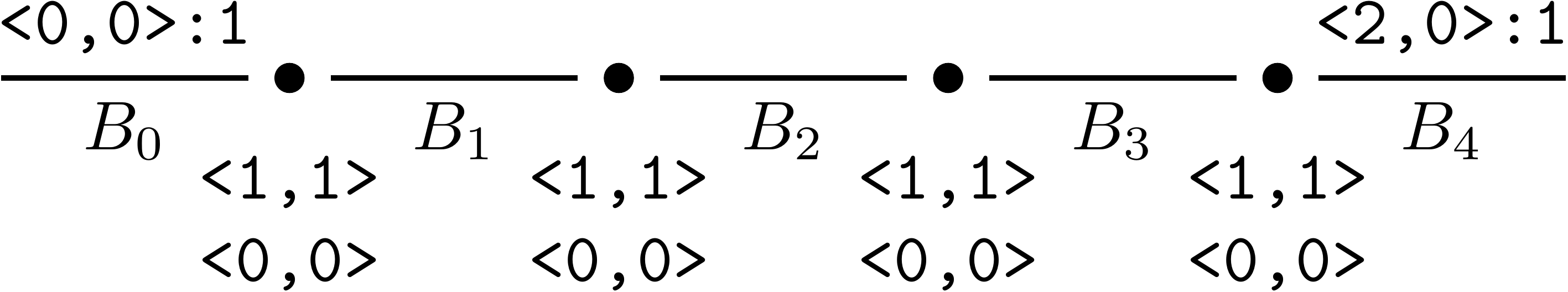}
\caption{Lattice comprised of four sites ($\bullet$) and five bonds $B$. Initial and target states, imposed at the incoming and outgoing bonds, 
are labeled according to Eq.~\eqref{eq:psilabel}. Physical indices corresponding to possible local states at the $l$-th site are given under each site.}
\label{fig:start_end_sectors}
\end{figure}

\begin{figure}[tbh]
\centering
		\includegraphics[scale=0.35]{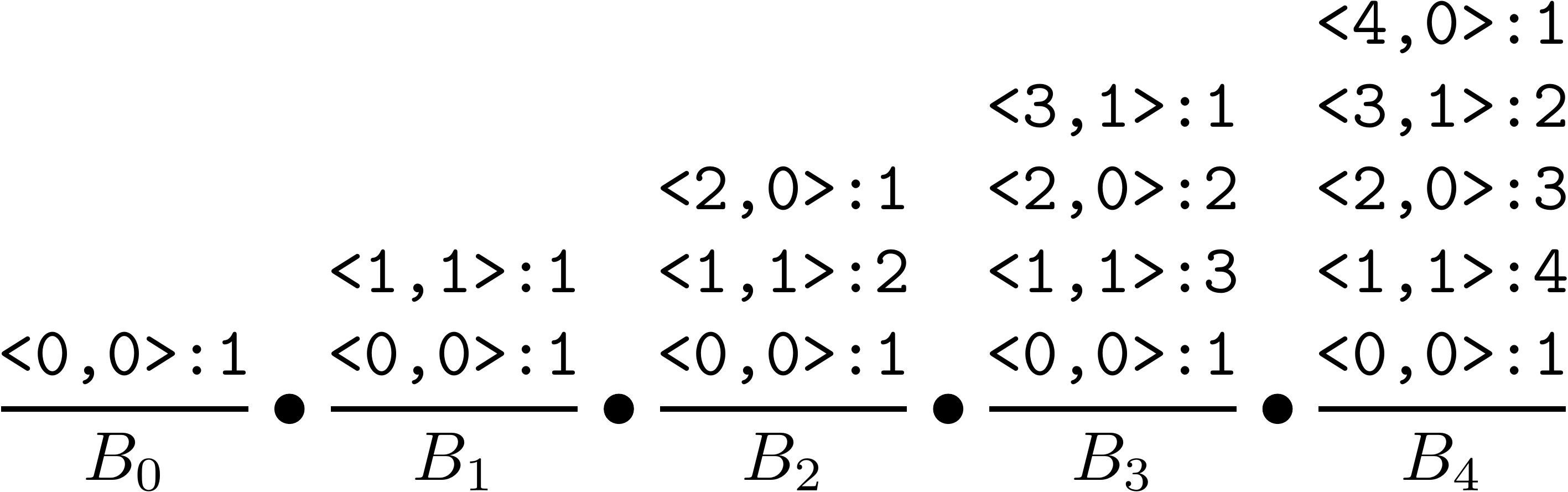}
	\caption{Possible symmetry sectors after traversing from left to right through the lattice. The charges on every bond are determined by combining the symmetry sectors on the previous bond with the on-site charges. The integer numbers paired with each charge count the occurrences of a given symmetry sector.}
	\label{fig:lr_sectors}
\end{figure}

\begin{figure}[tbh]
\centering
		\includegraphics[scale=0.35]{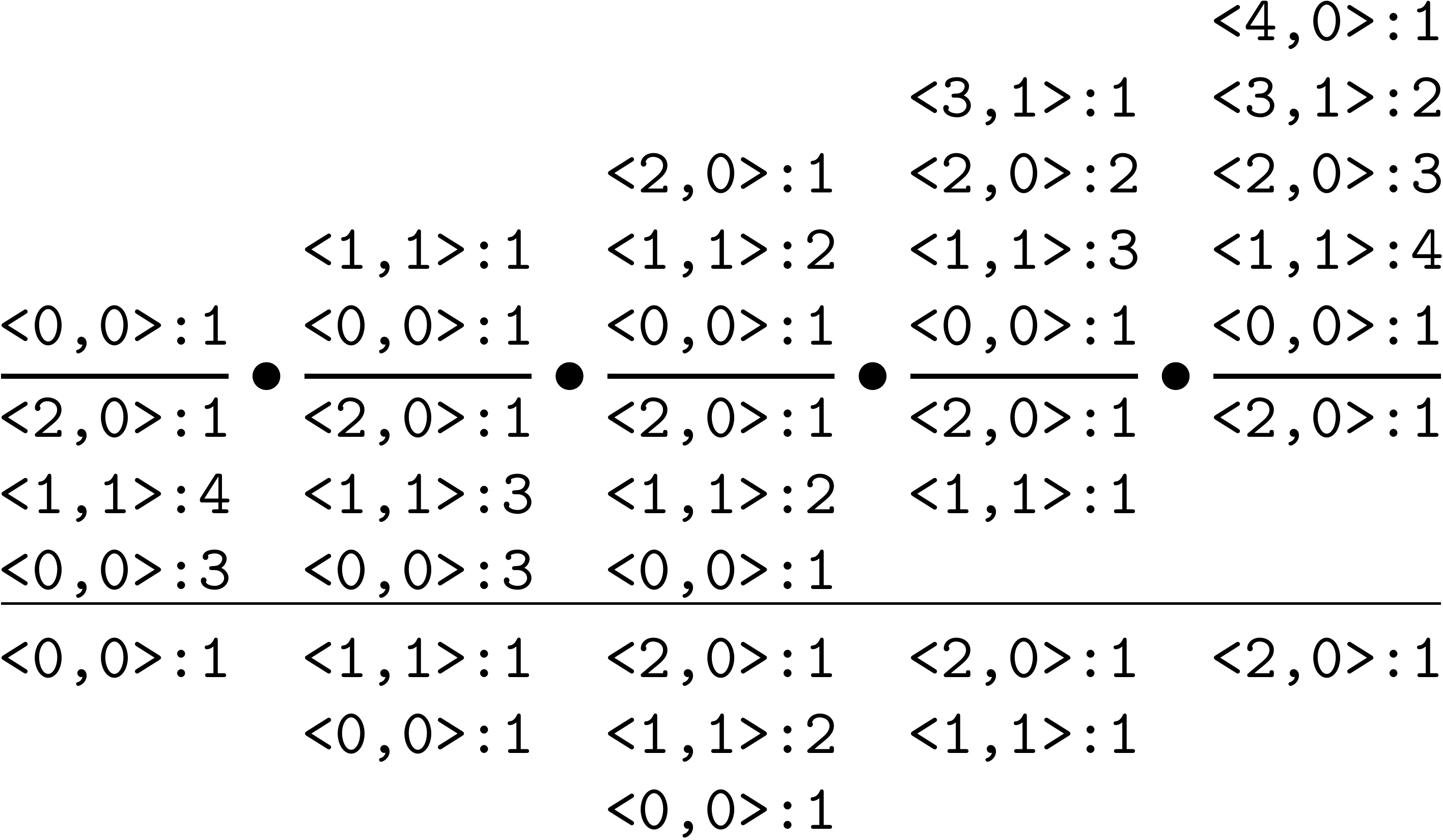}
	\caption{Illustration of the initial determination of a minimal subset of allowed (intermediate) charges as indicated on the last row. 
		The left and right boundary conditions are imposed by the target symmetry \texttt{g}\ and total particle number \texttt{N}\ 
		of a given problem. The indices --- shown as a pair of \texttt{charge:integer} on every bond --- are determined by combining the charges on 
		the previous bond with the on-site charges. The indices above (below) the lattice are generated by a left-to-right (right-to-left) traverse.}
	\label{fig:allowed_sectors}
\end{figure}

The indices are represented by sorted vectors of \texttt{charge:integer} pairs\ and are therefore uniquely defined with respect to the imposed symmetries. 
Hence, they constitute a placeholder for the quantum numbers and dimension of the blocks. In the initialization procedure, 
all possible charges are generated on each \textit{bond} of the lattice in a first step. 
To exemplify this procedure, we will in the following assume an active space of two electrons distributed over four spinors, 
e.g., a lattice of length $L=4$, as depicted in Figure \ref{fig:start_end_sectors}. 
Each site (corresponding to a spinor) is two-dimensional and can either be empty or occupied. 
The first number in \texttt{charge} represents the particle number and the second the double group irrep as follows from 
the definition of the left (right) index in Eq.~\eqref{eq:quant-num-for-MPS}. For simplicity, we further assume $\text{C}_{\text{1}}^{\ast}$ double group 
symmetry whose character and multiplication tables can be found in Table \ref{tab:c1charmult}. 
As can be seen from Table \ref{tab:c1charmult}, the double group $\text{C}_{\text{1}}^{\ast}$\ comprises 
two irreps $\Gamma_1$\ and $\Gamma_2$ which are symmetric (bosonic irrep) and antisymmetric (fermionic irrep), respectively, 
and are denoted in the following as 0 an $1$. 
Within this framework, we identify the initial state as the \emph{identity charge} given by \texttt{<0,0>}\ and the target state as 
\texttt{<2,0>}\ corresponding to a totally symmetric two-particle function. They are placed at left and right end of the lattice as 
shown in Figure \ref{fig:start_end_sectors}.
\begin{table}[tbh]
	\centering
	\caption{$\text{C}_{\text{1}}^{\ast}$\ character (left-hand side) and multiplication table (right-hand side) following the notation of Koster \textit{et al.} \cite{kost63}.}
	\begin{tabular}{l|rr}
		 $\text{C}_{\text{1}}^{\ast}$& E& $\bar{\rm E}$ \\
		\hline
		$\Gamma_1$ & 1 & 1\\
		$\Gamma_2$ & 1& -1\\
	\end{tabular}
	\hspace{1cm}
		\begin{tabular}{ll|r}
			 $\Gamma_1$ & $\Gamma_2$ & \\
			\hline
			$\Gamma_1$ & $\Gamma_2$ & $\Gamma_1$ \\
			& $\Gamma_1$ & $\Gamma_2$\\
		\end{tabular}
	\label{tab:c1charmult}
\end{table}
To proceed, we then need to identify all possible intermediate states of the system which is a key element to determine the actual block 
dimensions. By traversing the lattice from left to right, we obtain all possible charges at each bond by combining the ingoing charges --- corresponding to the outgoing charges on the previous bond --- with the local charges originating from the physical index at a given site $l$. 
This procedure will determine all possible states, also those which do not satisfy the symmetry constraints, as can be seen from 
Figure \ref{fig:lr_sectors}. The integer number associated to a given charge corresponds to the number of possible ways this 
charge can be realized and therefore determines the size of the associated block matrix in the MPS tensor (assuming that no further 
truncation in the subsequent MPS wave function optimization occurs). 
For instance, on bond $B_1$ \texttt{<1,1>} can only be realized by combining \texttt{<0,0>} on $B_0$ with the physical 
charge \texttt{<1,1>} on site 1. On bond $B_2$, \texttt{<1,1>}\ can be formed in two ways that is by combining either 
the outgoing charge \texttt{<1,1>} on $B_1$ with the local state \texttt{<0,0>} or by combining \texttt{<0,0>} with the local 
charge \texttt{<1,1>}. Following this principle leads to all states conserving the local symmetry on every bond. However, 
this is not sufficient to satisfy the overall symmetry of the total system.
For this purpose, we need to perform the same steps as before but now starting from the last bond B$_4$\ with the imposed target symmetry sector \texttt{<2,0>:1} and traverse the lattice from right to left. Moreover, going backwards necessitates to apply the inverse operations, for example,  
instead of directly employing the local charges, we first need to invert them before applying them to the ingoing charge. 
Having calculated all possible charge sectors and their dimensionality in both ways, 
the final step comprises the identification of all allowed states of the system. This requires to take the minimal subset of the charges identified 
in going from left-to-right and vice versa: only charges present in both directions are retained and the minimum dimension of each charge equivalent 
between the two directions is stored. The final result is illustrated in the lower part of Figure \ref{fig:allowed_sectors}. What we have just achieved is 
the ``pedestrian'' way to determine all possible intermediate states of the system which satisfy the imposed symmetry constraints both 
locally and globally. This allows us to illustrate the structure of a left-paired MPS\ consisting of left-paired MPS tensors (cf.~Figure \ref{fig:left_right_paired}). Left-pairing implies that the right basis of the matrix is equal to the right space while the left basis corresponds 
to the left and physical indices fused together. The ``construction recipe" for the rank-two tensor at site $l$, comprising a block-diagonal matrix,  
can be summarized as follows:
\begin{enumerate}
\item The number of blocks in the matrix is equal to the number of charges in the right basis.
\item The number of columns of each block corresponds to the dimension of the charge in the right basis.
\item The number of rows of each block corresponds to the sum of dimensions associated with all charges in the left space which can lead to the outgoing charge of the block. 
\end{enumerate}
Similar rules can be found for obtaining a right-paired MPS consisting of $L$\ right-paired MPS tensors.

\begin{figure}[tbh]
\centering
\includegraphics[scale=0.30]{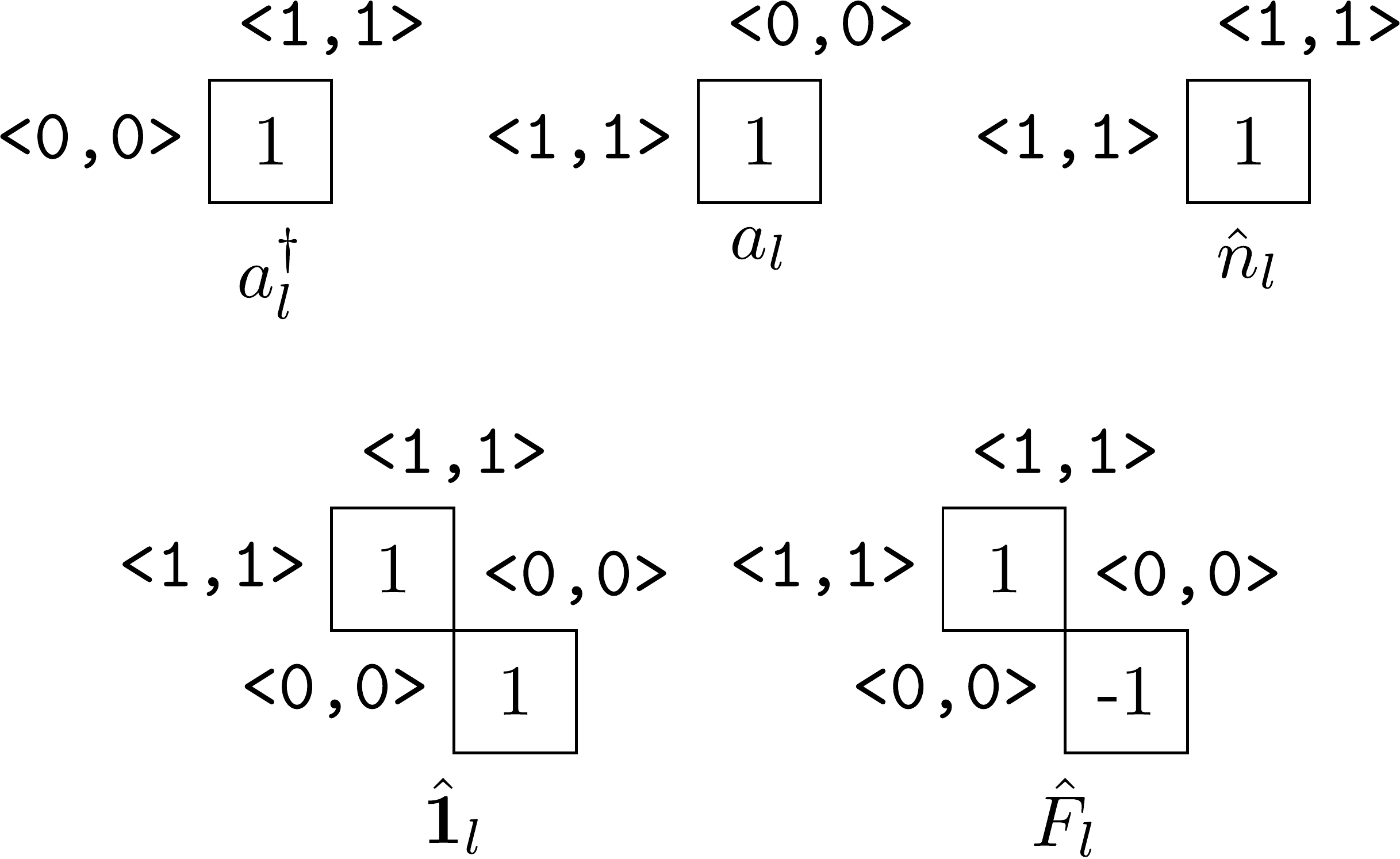}
\caption{Operators of the relativistic model as implemented in \qcm\ for the simple double group $\text{C}_{\text{1}}^\ast$.}
\label{fig:ops_rel_model}
\end{figure}

MPO tensors are defined analogously to the MPS ones. To illustrate their structure, we stay with the above example. 
Every site comprises a two-dimensional space which represents the spinor being either empty or occupied, corresponding to the local states \texttt{<0,0>} and \texttt{<1,1>}, respectively. A creation operator defined in such a basis brings us from an empty spinor 
to an occupied one and the associated block matrix is a one-dimensional matrix with ingoing charge \texttt{<0,0>} and outgoing charge \texttt{<1,1>}. Similarly, an annihilation operator will swap these charges leading from an occupied state to an empty one. 
In Figure \ref{fig:ops_rel_model}, we show the five elementary operators defined in the relativistic model associated with the double group C$_1^\ast$. The \texttt{MPO} class therefore defines the $\left\{{W}^{l}_{b_{l-1} b_{l}}\right\}$ matrices 
introduced in Eq.~(\ref{eq:mpoMOD}) containing the operators acting on each site $l$\ as depicted in Figure \ref{fig:ops_rel_model}\ 
for the C$_1^\ast$\ double group. Furthermore, in our relativistic model we also take advantage of the 
efficient MPO tensor-construction scheme discussed in detail for the nonrelativistic model in Ref.~\citenum{kell15a}. 

Turning next to the \texttt{Lattice} class shown in Figure \ref{fig:class_diagram_logic}, this class collects the type of each site, 
which, combined with the \texttt{Model} object, define the Hamiltonian and orbital (or spinor) basis. 
In our relativistic model, we implemented a spinor lattice for the second-quantized Hamiltonian (cf.~Eq.~(\ref{hamilrel})) which 
could in future even allow for working in a Kramers-unrestricted basis. The chosen lattice structure implies that 
it accommodates both unbarred and barred spinors while time-reversal symmetry is encoded by 
taking into account the appropriate block structure of the operator matrices (cf.~Section \ref{sec:rel-ham}).   
Finally, the \texttt{Sim} class shown in Figure \ref{fig:class_diagram_logic}\ 
constitutes the control center of \qcm, keeping, for example, pointers to \texttt{Lattice}\ and \texttt{Model}\ classes. The derived classes \texttt{dmrg\_sim} and \texttt{dmrg\_measure} 
drive the MPS optimization (see Section \ref{sec:mps-mpo-concepts-varopt})\ and measurement tasks such as the calculation of reduced density matrices (see Section \ref{sec:mps-rel-model-mpo-prop}), respectively. Further details concerning the full implementation of all classes illustrated in Figure \ref{fig:class_diagram_logic}\ 
can be found in Ref.~\citenum{kellerdiss}. 

\subsection{Molecular properties from relativistic DMRG wave functions}\label{sec:mps-rel-model-mpo-prop}

In the present work, we will primarily focus on molecular properties that are of expectation value type for a (Hermitian) one-particle operator 
$\hat{\Omega}$,
\beq
\left<\hat{\Omega}\right> = \text{tr}({\textbf{D}}^{s,\dagger} \boldsymbol{\Omega})\ ,
\label{eq:trace-exp}
\eeq
where 'tr' denotes the trace of a matrix and the bookkeeping index $s=\pm$\ indicates the properties of $\hat{\Omega}$ under time-reversal symmetry, with $s=+$ ($s=-$) being (anti-)symmetric (anti-symmetric), that is,
\beq
{{\hat{\mathcal K}}} {\hat{\Omega}} {{\hat{\mathcal K}}}^{-1} =
s{\hat{\Omega}}\ .
\label{kramers:timedef}
\eeq 
The elements $\{{D}^{s}_{pq}\}$ of the Kramers-restricted time-reversal (anti-)symmetric one-particle 
reduced density matrix (1-RDM) ${\textbf{D}}^{s}$ can be written in compact form with the help of the Kramers single-replacement operators (cf.~Eq.~\eqref{eq:kramx-one}), 
\beq
{D}^{s}_{pq} = \left<\Psi\left|\hat{X}^{s}_{pq}\right|\Psi\right>  
\qquad {D}^{s}_{\bar{p}q} = \left<\Psi\left|\hat{X}^{s}_{\bar{p}q}\right|\Psi\right> 
\qquad {D}^{s}_{{p}\bar{q}} = \left<\Psi\left|\hat{X}^{s}_{p\bar{q}}\right|\Psi\right>\ ,
\label{eq:1-rdm}
\eeq
which bears resemblance to the elements of a spin-traced 1-RDM in nonrelativistic theory.  
In passing, we note that similar expressions can be derived for (symmetrized) one-particle reduced transition density matrices  
by considering different bra and ket wave functions (of the same symmetry) $\bPsi$\ and $\kPhi$, respectively. 

Turning next calculation of property densities in a real space approach, requires as input a 1-RDM in atomic spinor basis rather than in molecular basis as obtained above \cite{bast_PhD} (in particular Chapter 5 and 
Appendix B of the cited work). 
For this purpose, we first back-transform the 
1-RDM matrix $\textbf{D}$\ comprising 
the elements given in Eq.\eqref{eq:1-rdm}\ 
from molecular to atomic spinor basis by means of the molecular spinor coefficients $\{C_{\lambda p}\}$, 
\beq
\mathcal{D}^{s}_{} = \sum\limits_{\lambda\kappa} C_{\lambda p} D^s_{pq} C_{\kappa q}^{\ast}\ .  
\label{eq:back-1rdm}
\eeq
In a second step, if required, 
we construct, a time-reversal symmetric atomic spinor density matrix $\mathcal{D}^{+}$ by extracting the imaginary phase from the corresponding time-reversal anti-symmetric matrix where the elements of the symmetric matrix (dropping the upper index $+$\ from now on) read as \cite{bast_PhD},
\beq
{\mathcal{D}} =
\left(\begin{array}{cc}
D_{\lambda\kappa}& D_{\lambda\bar{\kappa}} \\
D_{\bar{\lambda}\kappa} &  D_{\bar{\lambda}\bar{\kappa}} \end{array}\right)\ .
\label{kramers:opbasD}
\eeq
The property density calculation in a Kramers-restricted atomic spinor basis then proceeds from the general \textit{ansatz} \cite{bast_PhD},
\beq\label{eq:dens-general}
\Psi^\dagger \hat{\Omega} \Psi = f \Psi^{\rm L\dagger} {\Omega}^{\rm LX} \Psi^{\rm X} + g \Psi^{\rm S\dagger} {\Omega}^{\rm SY} \Psi^{\rm Y}\ ,
\eeq
where $\Psi$\ corresponds to our four-component (optimized) MPS wave function with L representing its large and S the corresponding 
small component and (X,Y)\ being either (L,S)\ or (S,L). Furthermore, $f$\ and $g$\ are scalar (complex) factors while the one-electron operator 
$\hat{\Omega}$\ represents a specific Dirac matrix as discussed in the following. 
In case of the current density \cite{dyal07,reih09}
\beq\label{eq:j}
\textbf{j} = -e \Psi^{\dagger} c\boldsymbol{\alpha} \Psi\ ,
\eeq
and magnetization \cite{dyal07,reih09}
\beq\label{eq:m}
\textbf{m} = \Psi^{\dagger} \boldsymbol{\Sigma} \Psi\ , 
\eeq
these operators read as $\hat{\Omega}=i\boldsymbol{\alpha}$\ and $\hat{\Omega}=i\boldsymbol{\Sigma}$, respectively. The imaginary phase $i$\ is required since both the Dirac $\boldsymbol{\alpha}$\ matrices, given here in $x,y,z$-direction as
\beq
 \boldsymbol{\alpha}=\left[\boldsymbol{\alpha}_x,\boldsymbol{\alpha}_y,\boldsymbol{\alpha}_z\right] = 
 \left[
 \left(\begin{array}{cccc}
 \hfill 0 \hfill & \hfill 0 \hfill & \hfill 0 \hfill & \hfill 1 \hfill \\
  \hfill 0 \hfill & \hfill 0 \hfill & \hfill 1 \hfill & \hfill 0 \hfill \\
  \hfill 0 \hfill & \hfill 1 \hfill & \hfill 0 \hfill & \hfill 0 \hfill \\
  \hfill 1 \hfill & \hfill 0 \hfill & \hfill 0 \hfill & \hfill 0 \hfill  \end{array} \right),
  \left(\begin{array}{cccc}
   \hfill 0 \hfill & \hfill 0 \hfill & \hfill 0 \hfill & \hfill -i \hfill \\
   \hfill 0 \hfill & \hfill 0 \hfill & \hfill i \hfill & \hfill 0 \hfill \\
   \hfill 0 \hfill & \hfill -i \hfill & \hfill 0 \hfill & \hfill 0 \hfill \\
   \hfill i \hfill & \hfill 0 \hfill & \hfill 0 \hfill & \hfill 0 \hfill  \end{array} \right),
   \left(\begin{array}{cccc}
    \hfill 0 \hfill & \hfill 0 \hfill & \hfill 1 \hfill & \hfill 0 \hfill \\
    \hfill 0 \hfill & \hfill 0 \hfill & \hfill 0 \hfill & \hfill -1 \hfill \\
    \hfill 1 \hfill & \hfill 0 \hfill & \hfill 0 \hfill & \hfill 0 \hfill \\
    \hfill 0 \hfill & \hfill -1 \hfill & \hfill 0 \hfill & \hfill 0 \hfill  \end{array} \right)
 \right]\ , 
 \eeq
 and the corresponding Dirac $\boldsymbol{\Sigma}$\ matrices, given here in $x,y,z$-direction as 
\beq
\boldsymbol{\Sigma} = \left[\boldsymbol{\Sigma}_x,\boldsymbol{\Sigma}_y,\boldsymbol{\Sigma}_z\right] = 
\left[
 \left(\begin{array}{cccc}
 \hfill 0 \hfill & \hfill 1 \hfill & \hfill 0 \hfill & \hfill 0 \hfill \\
  \hfill 1 \hfill & \hfill 0 \hfill & \hfill 0 \hfill & \hfill 0 \hfill \\
  \hfill 0 \hfill & \hfill 0 \hfill & \hfill 0 \hfill & \hfill 1 \hfill \\
  \hfill 0 \hfill & \hfill 0 \hfill & \hfill 1 \hfill & \hfill 0 \hfill  \end{array} \right),
  \left(\begin{array}{cccc}
   \hfill 0 \hfill & \hfill -i \hfill & \hfill 0 \hfill & \hfill 0 \hfill \\
   \hfill i \hfill & \hfill 0 \hfill & \hfill 0 \hfill & \hfill 0 \hfill \\
   \hfill 0 \hfill & \hfill 0 \hfill & \hfill 0 \hfill & \hfill -i \hfill \\
   \hfill 0 \hfill & \hfill 0 \hfill & \hfill i \hfill & \hfill 0 \hfill  \end{array} \right),
   \left(\begin{array}{cccc}
    \hfill 1 \hfill & \hfill 0 \hfill & \hfill 0 \hfill & \hfill 0 \hfill \\
    \hfill 0 \hfill & \hfill -1 \hfill & \hfill 0 \hfill & \hfill 0 \hfill \\
    \hfill 0 \hfill & \hfill 0 \hfill & \hfill 1 \hfill & \hfill 0 \hfill \\
    \hfill 0 \hfill & \hfill 0 \hfill & \hfill 0 \hfill & \hfill -1 \hfill  \end{array} \right)
\right]\ ,
\eeq
are time-reversal antisymmetric \cite{dyal07,reih09}. 
Note that other definitions of \textbf{m}\ exist, originating from a Gordon decomposition of the four-component relativistic charge current density \cite{gord28b,saku67a,jaco12a}, but we follow in the present work the definition of Ref.~\citenum{bast_PhD}\ which states that $\boldsymbol{\Sigma}$\ is the ``natural" 
relativistic analogue for the (nonrelativistic) spin-operator $\boldsymbol{\sigma}$, e.g., the Pauli spin matrices, 
in a four-component relativistic framework. 

By virtue of Eq.~\eqref{kramers:opbasD}, the first term on the right-hand side of Eq.~\eqref{eq:dens-general}\ can be written as \cite{bast_PhD},
\beq
f \Psi^{\rm L\dagger} {\Omega}^{\rm LX} \Psi^{\rm X} = f S_{\lambda\kappa}^{\rm LX} \left[
\Omega_{\alpha\alpha}^{\rm LX}D_{\lambda\kappa}^{\rm XL}+
\Omega_{\alpha\beta}^{\rm LX}D_{\lambda\bar{\kappa}}^{\rm XL}+
\Omega_{\beta\alpha}^{\rm LX}D_{\bar{\lambda}\kappa}^{\rm XL}+
\Omega_{\beta\beta}^{\rm LX}D_{\bar{\lambda}\bar{\kappa}}^{\rm XL}
\right]\ ,
\eeq
where $S_{\lambda\kappa}^{\rm LX}$\ is the atomic spinor overlap distribution $S_{\lambda\kappa}^{\rm LX}=\left<\varphi_{\lambda}^{\rm L}\left|\right.\varphi_{\kappa}^{\rm X}\right>$. Likewise, similar considerations hold for the second term on the right-hand side of Eq.~\eqref{eq:dens-general}. In Table \ref{tab:props-contrib}\ we summarize 
the contributions needed for the evaluation of the property densities considered in this work, namely the current density \textbf{j}\ and magnetization \textbf{m}. 
\begin{table}[tbh]
\caption{\label{tab:props-contrib}Quantities needed to evaluate property densities $\Psi^\dagger \hat{\Omega} \Psi$\ according to Eq.~\eqref{eq:dens-general}.  
This table has been adopted from Table B.1 in the Appendix of  Ref.~\citenum{bast_PhD}.}
\centering
\begin{tabular}{lcrrrr}\hline\hline
$\hat{\Omega}$ & $D$-block & $f$ & $g$ & X & Y\\ \hline
$i\boldsymbol{\Sigma}_x$ & $\bar{\lambda}\kappa$ & $i$ & $i$ & L & S \\
$i\boldsymbol{\Sigma}_y$ & $\lambda\bar{\kappa}$ & $-1$ & $-1$ & L & S \\
$i\boldsymbol{\Sigma}_z$ & $\bar{\lambda}\bar{\kappa}$ & $i$ & $i$ & L & S \\
$i\boldsymbol{\alpha}_x$ & $\bar{\lambda}\kappa$ & $i$ & $i$ & S & L \\
$i\boldsymbol{\alpha}_y$ & $\lambda\bar{\kappa}$ & $-1$ & $-1$ & S & L \\
$i\boldsymbol{\alpha}_z$ & $\bar{\lambda}\bar{\kappa}$ & $i$ & $i$ & S & L \\
\hline
\end{tabular}
\end{table}

\section{Numerical examples}\label{sec:mps-rel-numerical}

\subsection{Computational details}\label{sec:mps-rel-comp-det}

To benchmark our relativistic DMRG implementation in \qcm,  
we performed single-point energy calculations for the TlH molecule 
employing the same computational setup as described in Ref.~\citenum{knec14}. 
A short summary of the computational details is given in the following. 
Orbitals and integrals in molecular spinor basis were computed  
with a development version of the \dir\ program package \cite{DIRAC12} using the four-component 
Dirac-Coulomb Hamiltonian and triple-$\zeta$\ basis sets for Tl (\texttt{dyall.cv3z}) \cite{dyal02,dyal12a} and H (cc-pVTZ) \cite{dunn89}. 
C$_{32v}^{\ast}$\ double group symmetry was assumed 
throughout all calculations for TlH. The active space for the DMRG-CI step comprised 
14 electrons --- corresponding to the occupied Tl $5d6s6p$ plus H $1s$\ shells --- in 47 Kramers pairs (94 spinors). 
The DMRG-CI calculations are mainly characterized by two parameters, (i) the number of renormalized states $m$\ and (ii) 
the truncation tolerance $\delta$ of the singular value spectrum. 
For the latter, we employed two different values in all calculations, namely 
an initial tolerance of $\delta_{\rm initial}$ =10$^{-40}$ and a final one $\delta_{\rm final}$=10$^{-9}$. 
In the first three sweeps of each calculation we log-interpolated the truncation from $\delta_{\rm initial}$\ to $\delta_{\rm final}$, 
which was then applied until convergence. Subsequently, the $m$\ value was increased after each one of the first ten sweeps and 
then kept constant until the end of the optimization. The data are denoted as DMRG-CI[$m$]\ where [$m$]\ denotes the final number of renormalized states $m$.  
Other important factors determining accuracy and convergence rate are the chosen initial state and spinor order \cite{kell14}. 
We employed a single-determinant guess for the environment states in the initial sweep with the determinant corresponding to the Hartree-Fock solution. 
Moreover, the spinors were ordered according to the natural orbital occupation numbers obtained 
from a preceding perturbation theory calculation (see Ref.~\citenum{knec14}) 
and by pairing Kramers partners, i.e., placing barred spinors next to their unbarred, time-reversal partners.

\begin{figure}[tbph]
\centering
  \includegraphics[scale=0.35]{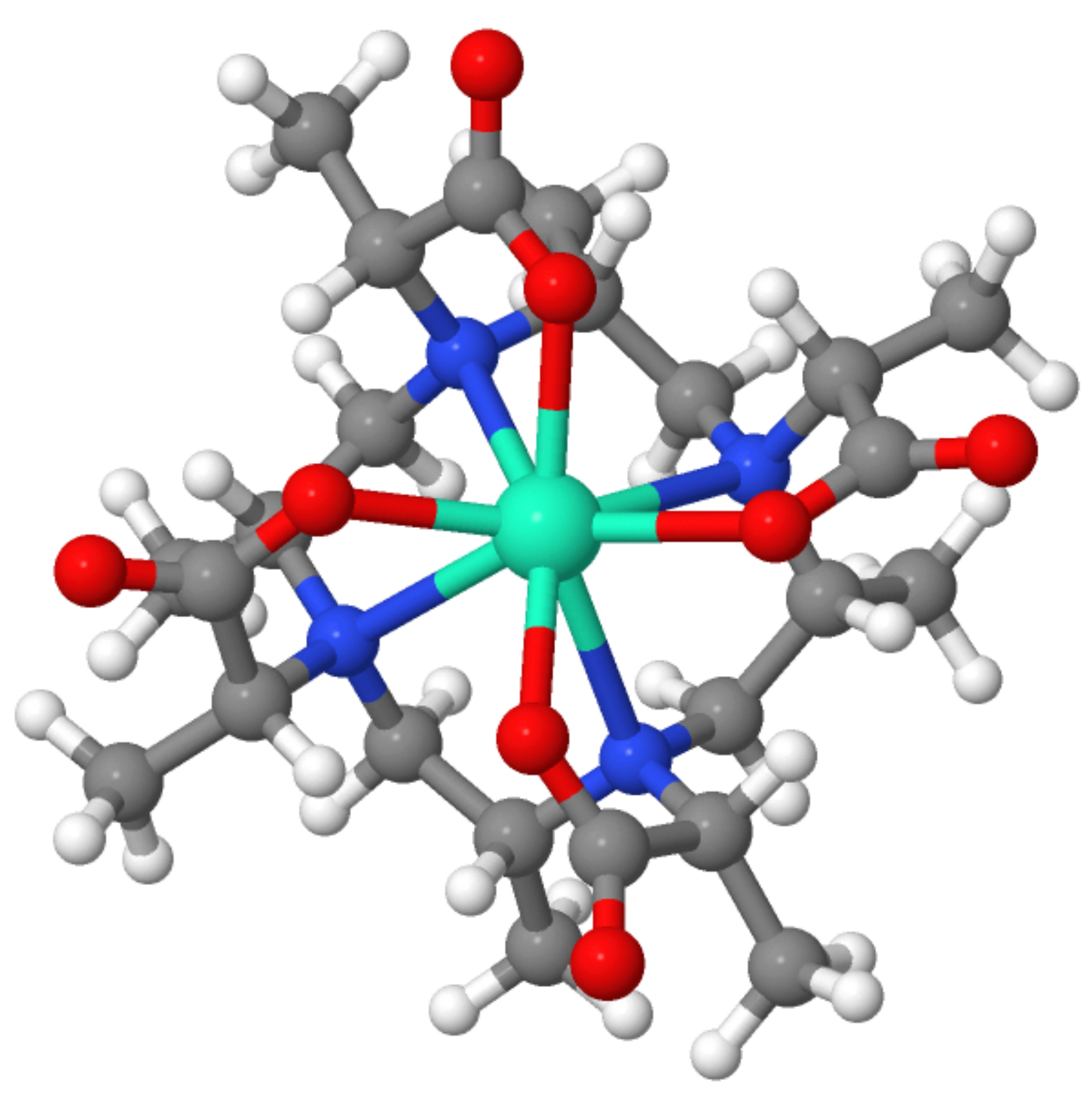}
\caption{Optimized (B3LYP/ECP(Dy)/TZVP) structure of the \dyc\ complex.}
\label{fig:dy_complex}
\end{figure}
The structure of the C$_4$-symmetric \dyc\ complex shown in Figure \ref{fig:dy_complex}\ was 
optimized with density functional theory (DFT)  by means of the B3LYP functional \cite{beck93a,leec88} as implemented in Gaussian09 \cite{g09}. 
A Stuttgart/Dresden effective core potential with 55 core electrons \cite{dolg93} along with the corresponding 
valence TZVP \cite{scha94} basis set was employed for the Dy atom and, accordingly, a TZVP basis for the ligand atoms O, H, C and N.
Subsequent four-component calculations were carried out within the framework of the \dir\ program \cite{DIRAC12}\ employing (Abelian) C$_2^\ast$\ double group symmetry. 
We obtained the starting Kramers pair basis for the DMRG-CI calculations from an average-of-configuration Hartree-Fock 
calculation with nine electrons distributed in 14 spinors (corresponding to the $f^9$\ open valence shell of Dy(III)) 
based on the four-component Dirac-Coulomb Hamiltonian. 
Uncontracted large-component basis sets of double-$\zeta$\ quality (dubbed as \texttt{DZ}) were 
employed for Dy (dyall.v2z) \cite{gome10b} as well as for the ligand atoms (cc-pVDZ basis set \cite{dunn89}). 
The small component basis set was generated using restricted kinetic balance. 
Starting from the optimized Hartree-Fock orbitals, we chose three different active orbital spaces for the subsequent DMRG-CI 
calculations, namely (i) a CAS(9,14) comprising all Dy $4f$\ valence electrons and spinors, (ii) a CAS(27,32) adding the subvalence Dy 
$4p4s4d$\ shell to the previous CAS and (iii) CAS(27,58) augmenting the preceding CAS with secondary spinors that 
have significant contributions from both Dy and ligand orbitals as determined from a Mulliken population analysis of the Hartree-Fock wave function. 
The current density and magnetization were obtained within \texttt{VISUAL}\ module of \dir\ and 
visualized by means of the \textsc{pyngl-streamline} program \cite{pyngl}. To this end, the \texttt{VISUAL}\ module had 
been extended to read a one-particle reduced density matrix in molecular spinor basis provided by \qcm\ which was 
subsequently transformed from molecular to atomic spinor basis to yield the density matrix \textbf{D} (cf.~Eq.\eqref{kramers:opbasD})\ that constituted the starting point for the property density calculation. 
The DMRG-CI wave function optimizations and property density calculations employed a common gauge origin, 
placed at the center of mass. 
The visual integration was carried out within the respective planes specified in Section \ref{sec:mps-rel-dy} by employing a 
$12$ bohr $\times$ $12$ bohr grid with the Dy(III) center placed at the (0,0,0) origin and 600 steps along each side corresponding to $600*600=360000$\ integration points. Similar to the DMRG-CI setup for TlH, spinors for the MPS optimization 
were ordered according to orbital energies and pairing Kramers partners. 
The initial guess for the MPS in the starting sweep was based on random numbers (option \texttt{init\_state=default}\ in \qcm). 
The sweep procedure was terminated after converge to 10$^{-6}$\ Hartree was reached for a given number of renormalized states $m$.

\subsection{TlH}\label{sec:mps-rel-tlh}

In order to assess the capabilities of our new four-component DMRG implementation, we compare absolute energies for the electronic $\Omega=0^+$\ ground state 
of the thallium hydride molecule at different Tl-H internuclear distances to our benchmark data provided in Ref.~\citenum{knec14}. 

\begin{figure}[tbh]
	\centering
	\includegraphics[scale=0.40]{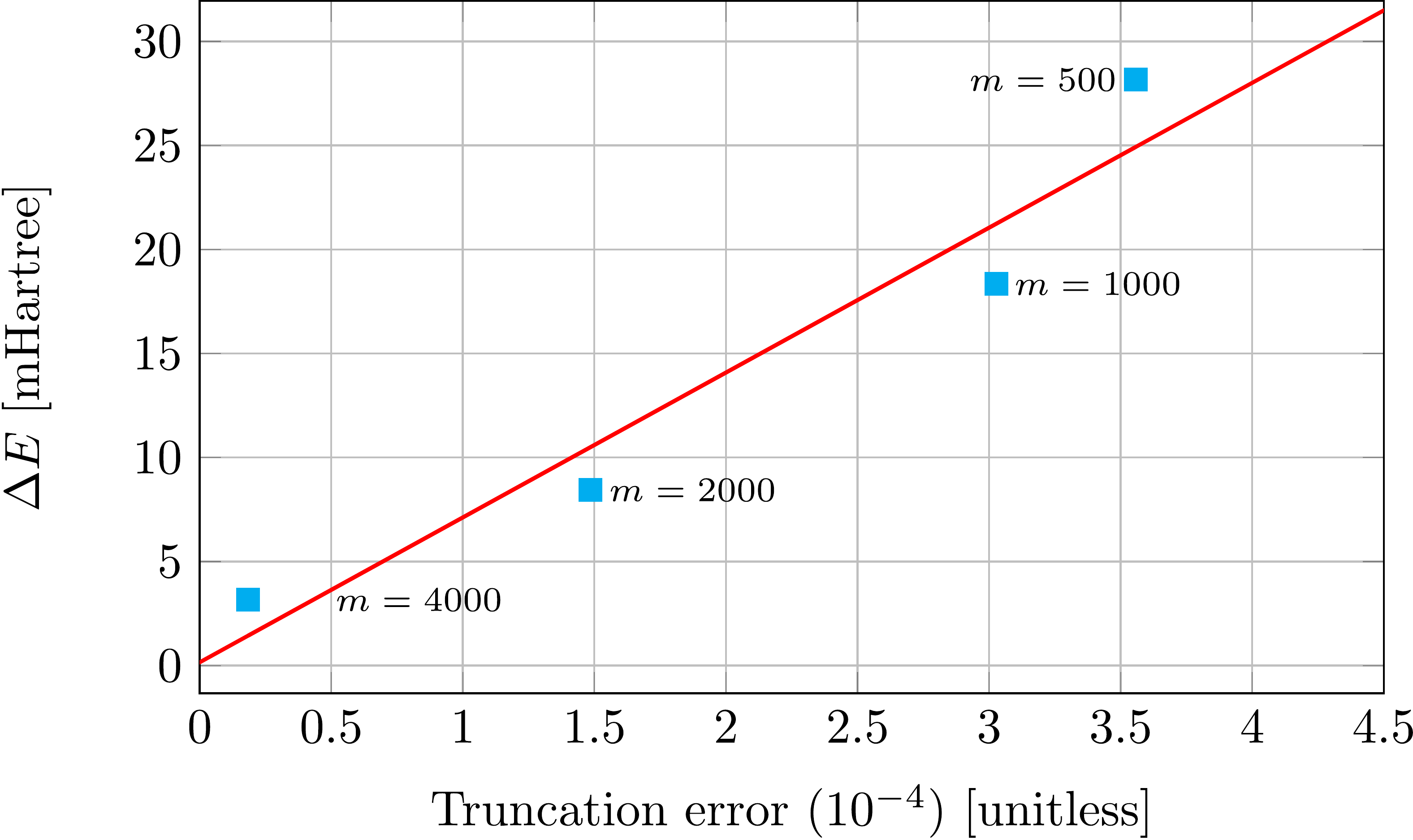}
	\caption{Extrapolation $\Delta E_{}$ for $m\rightarrow \infty$\ for the data obtained with \qcm. The difference to the reference CCSDTQ energy for the final extrapolated data is $0.1$ mHartree.}
	\label{fig:extrapolation}
\end{figure}

Inspection of Table \ref{tab:energies_TlH}\ reveals that the four-component DMRG-CI energy with $m = 5000$ is, 
although being below the best variational CI energy (4c-CISDTQ in Table \ref{tab:energies_TlH}), 
still higher than the reference four-component CCSDTQ energy. 
We recall that DMRG is best suited for static-correlation problems, whereas TlH (close to equilibrium) is dominated by dynamic correlation as discussed on the basis of entanglement measures in Ref.~\citenum{knec14}.  
In addition, the latter claim is supported by a recent study of Shiozaki and Mizukami \cite{shio15a} who considered 
ground-state spectroscopic properties of TlH based on a relativistic CASSCF reference wave function employing a minimal 
CAS(4,5) active space. They demonstrated that a subsequent treatment of dynamic-correlation contributions by means of multireference approaches such as CASPT2, NEVPT2 or internally-contracted MRCI+Q is sufficient to obtain excellent agreement with experimental reference data.  
Returning to our DMRG-CI results, Figure \ref{fig:extrapolation} illustrates the extrapolation of 
the DMRG-CI energy for an increasing number of renormalized states $m$ to the limit\ $E_{}(m\rightarrow\infty)$\ which 
provides an effective means to eliminate the truncation error. 
Following this procedure yields a best estimate of $\Delta E_{}(m\rightarrow\infty)$\ at R$_e$=1.872 \AA\  
as close as +0.1 mHartree to the four-component CCSDTQ reference energy.
\begin{table}[tbh]
\caption{Total electronic energy differences $\Delta E_{}$ (in mHartree) for different correlation approaches with respect to the four-component (4c-)CCSDTQ reference 
energies taken from Ref.~\citenum{knec14}. DMRG-CI results were obtained with an increasing number of renormalized block states $m$. 
For 4c-DMRG-CI[$m=5000$], energy differences are listed for five different internuclear distances $R_{\rm Tl-H}$ (in \AA) given as subscripts.}	
\centering
\begin{tabular}{l@{\hspace{5mm}}ccccc}\hline \hline 
  Method & $\Delta E_{1.7720}$ & $\Delta E_{1.8220}$ & $\Delta E_{1.8720}$ & $\Delta E_{1.9220}$ & $\Delta E_{1.9720}$ \\ 
  \hline
  4c-CISDTQ (Ref.~\citenum{knec14}) & 2.57 & 2.60 & 2.63 & 2.66 & 2.70 \\
  4c-CCSDT (Ref.~\citenum{knec14}) & 0.32 & 0.33 & 0.33 & 0.33 & 0.33 \\
  4c-CCSDT(Q) (Ref.~\citenum{knec14}) & -0.07 & -0.06 & -0.07 & -0.07 & -0.06 \\
  4c-DMRG-CI[$m=4500$] (Ref.~\citenum{knec14}) & - & - & 2.57 & - & - \\
  4c-DMRG-CI[$m=500$] & - & - & 28.16 & - & - \\
  4c-DMRG-CI[$m=1000$] & - & - & 18.35 & - & - \\
  4c-DMRG-CI[$m=2000$] & - & - & 8.44 & - & - \\
  4c-DMRG-CI[$m=4000$] & -& - & 3.16 & - &- \\
  4c-DMRG-CI[$m=5000$] & 2.53 & 2.49 & 2.51 & 2.45 & 2.51 \\
  4c-DMRG-CI[$m \rightarrow \infty$] & - & - & 0.1 & - & - \\ \hline
  \end{tabular}
\label{tab:energies_TlH}
\end{table}

\begin{figure}[tbh]
	\centering
		\includegraphics[scale=0.485]{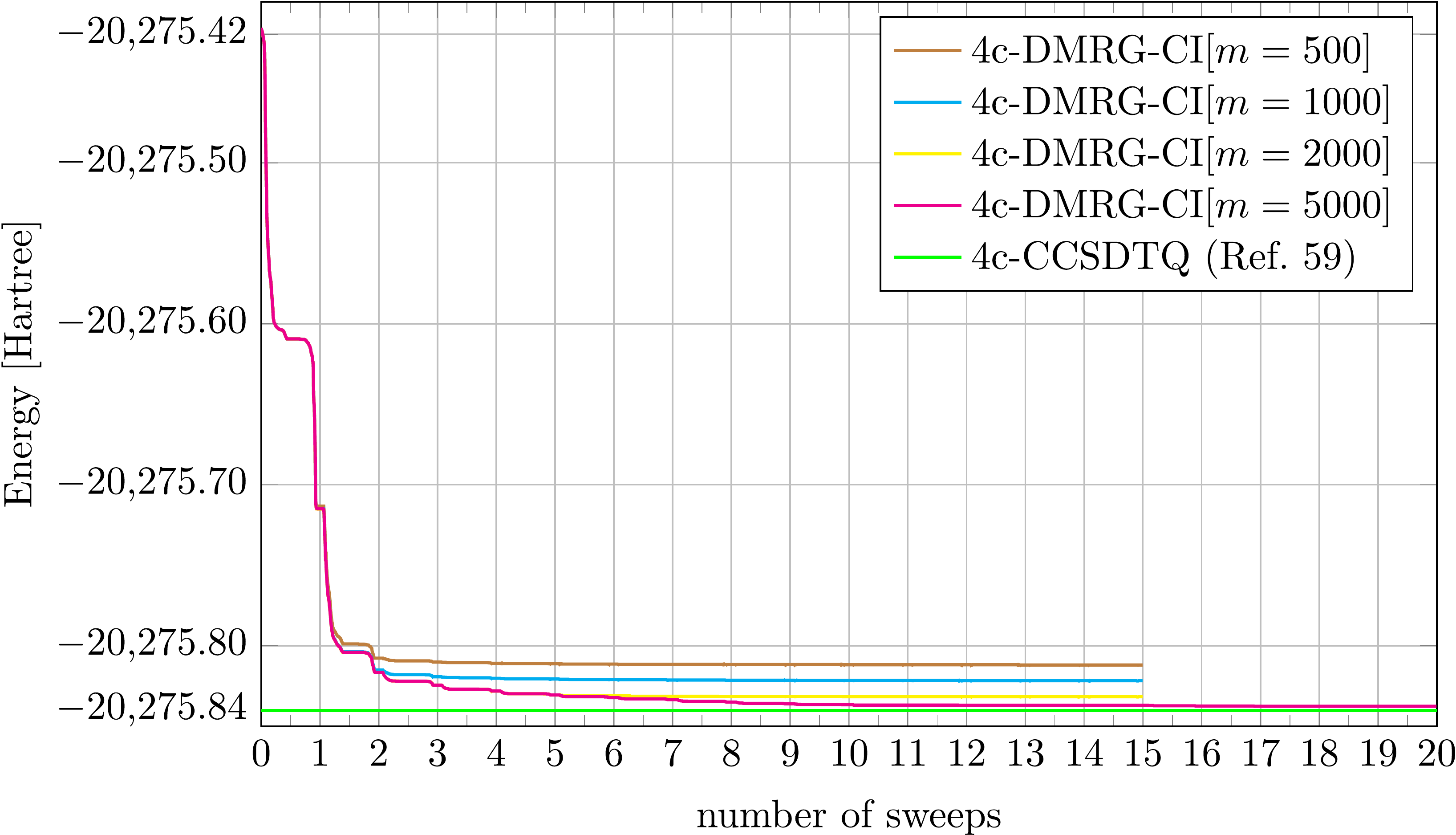}
	\caption{Energy convergence of a DMRG-CI calculation as a function of the number of sweeps performed 
	for different values of $m$ at the experimental equilibrium internuclear distance R$_e=1.872$ \AA\ of the TlH molecule. 
	For values of $m$\ lower than $5000$, the sweeping procedure was 
	terminated after fifteen iterations.}
	\label{fig:energy_conv}
\end{figure}

Finally, to achieve convergence in a DMRG-CI calculation usually requires a sufficient number of sweeps as can be seen in Figure \ref{fig:energy_conv}. 
For example, the results obtained with $m=5000$\ necessitated between 15 to 20 sweeps (depending on the internuclear Tl-H distance) to ensure a saturation of the (correlation) energy for the chosen $m$ value. Note that in the context of \qcm\ a total sweep  corresponds to a combined left-to-right \textit{and}\ right-to-left sweep as indicated in Figure \ref{fig:mps-orbital-1D}. 

\subsection{Magnetic properties}\label{sec:mps-rel-dy}

In the previous section, we assessed the performance of our relativistic DMRG approach for the calculation of total energies in 
a small diatomic molecule by comparison to data obtained with our pilot implementation \cite{knec14}. 
In this section, we illustrate the potential of our new relativistic DMRG by examining selected magnetic properties of a Dy(III)-containing complex which goes beyond earlier capabilities. Because of symmetry restrictions \footnote{A bug in the \dir\ software package prevented us from obtaining correct 
two-electron integrals in molecular spinor basis for the C$_1^\ast$\ double group. By means of the \textsc{Bagel}\ program package \cite{bagel}, we could rule out an inconsistency in our \qcm\ implementation for the C$_1^\ast$\ double group.}, we consider in the following 
a model Dy(III) complex with the chelating octa-methyl DOTA (1,4,7,10 tetraaza-cyclododecane-tetraacetic acid) ligand denoted \mmod\ and shown in Figure \ref{fig:dy_complex}\ in contrast to the eight-fold, stereospecifically methylated DOTA derivatives dubbed as \texttt{M8} that were employed 
in the work of H\"aussinger \mea \cite{haus09}. 
Owing to their high affinity towards (trivalent) lanthanides, DOTA complexes are ubiquitous in chelating lanthanides 
and are widely used for applications in magnetic resonance imaging \cite{merb13,faul14}, biomedical imaging \cite{bunz10,stas13}, biological nuclear magnetic 
resonance \cite{wuth86,pigu03,pint07,clor09a,bert17} and EPR-spectroscopy based distance measurements in proteins \cite{pann00,jesc07,jesc12,gold14}.

The chemistry of lanthanides is characterized in general --- although a rich chemistry of Ln(II) complexes does exist as well, see for example Ref.~\citenum{szos12} --- by the $+3$\ oxidation state omnipresent in the whole series as well as the compact nature of the $4f$\ orbitals 
that exhibit an almost core-like character \cite{cott06}. 
As a result, the $4f$\ electrons contribute only to a limited extent to the chemical bonding which, in turn, leads to no strong preferences in coordination numbers and coordination polyhedra in lanthanide complexes of specific $4f$-shell configurations. 
Still, magnetic properties and electronic transitions of Ln complexes are sensitive to the 
coordination environment, though to a lesser extent than SO coupling and interelectronic repulsion effects, 
which makes them interesting target compounds for applications in magnetism and luminescence \cite{cott06}.

We therefore chose in this work to study the (charge) current density $\textbf{j}$\ and magnetization $\textbf{m}$\ of the free Dy(III) ion in comparison with the \dyc\ complex shown in Figure \ref{fig:dy_complex}. For this purpose, we carried out  
four-component DMRG-CI calculations with our newly developed relativistic  DMRG model. 
By virtue of the expectation-value approach outlined in Section \ref{sec:mps-rel-model-mpo-prop}, we obtained current densities and magnetizations for the bare Dy$^{3+}$\ ion and the \dyc\ complex 
in a real-space approach based on DMRG-CI wave functions with active orbital spaces of increasing size. 
As discussed in Ref.~\citenum{bast_PhD}, a real-space approach  
allow us to visualize property densities without having to worry 
about orthonormality terms which would otherwise be needed in an analytical approach. 
\begin{figure}[tbhp]
\centering
  \includegraphics[scale=0.385]{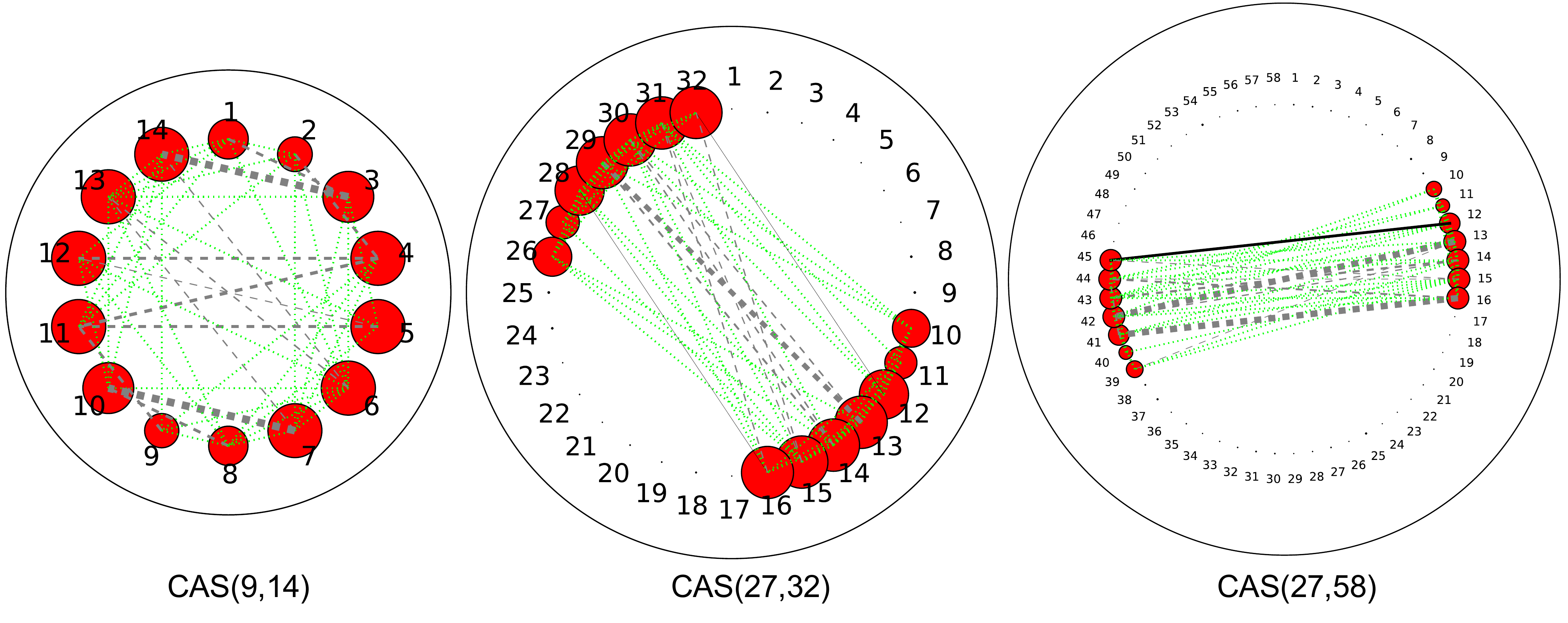}
\caption{Entanglement diagrams for the electronic ground state of the \dyc\ complex obtained from 
4c-DMRG-CI/\texttt{DZ} wave functions with increasing active orbital spaces. 
	The area of the red circles indicates the magnitude of the single-orbital entropy $s_p(1)$. The line connecting two spinors $p$\ and $q$\ denotes their mutual information 
	value $I_{pq}$ where darker (from green to black) and thicker lines correspond to an increasing value of the mutual information.}
\label{fig:entanglement_dycomplex_1}
\end{figure}

Figure \ref{fig:entanglement_dycomplex_1}\ summarizes the single-orbital entropy $s_p$(1) \cite{lege03a} for each spinor $p$ and the mutual information $I_{pq}$ \cite{riss06,lege06}\ for spinor pairs $p,q$\ calculated for the \dyc\ complex with increasing 
active orbital spaces. 
In all ring diagrams, spinors are ordered with all unbarred spinors first followed by their time-reversal partners. The diagram representation follows the scheme defined in Ref.~\citenum{stei16a}: the area of the red circles assigned to each spinor $p$ is proportional to its single-orbital entropy $s_p(1)$\ 
whereas the line connecting two spinors $p,q$\ denotes their mutual information value $I_{pq}$. The direct connection between (spinor) entanglement and non-dynamical electron correlation --- originally examined for spatial orbitals in a nonrelativistic framework \cite{bogu12a} --- 
provides a suitable means to analyze the active space in terms of non-dynamical electron correlation contributions originating  
from spinors exhibiting different character and/or interactions.

The entanglement diagrams in Figure \ref{fig:entanglement_dycomplex_1}\  unequivocally illustrate that the valence 
electronic structure of the \dyc\ complex is predominantly 
governed by electron correlation within the partially filled $4f$\ shell. 
For example, extending the active space beyond a CAS(9,14), corresponding to correlating all nine $4f$\ electrons within the 14 $4f$\ spinors, 
has virtually no effect neither on the single-orbital entropies $s_p(1)$\ nor on the magnitude of the mutual information $I_{pq}$. 

\begin{figure}[tbh]
\centering
	\includegraphics[scale=0.515]{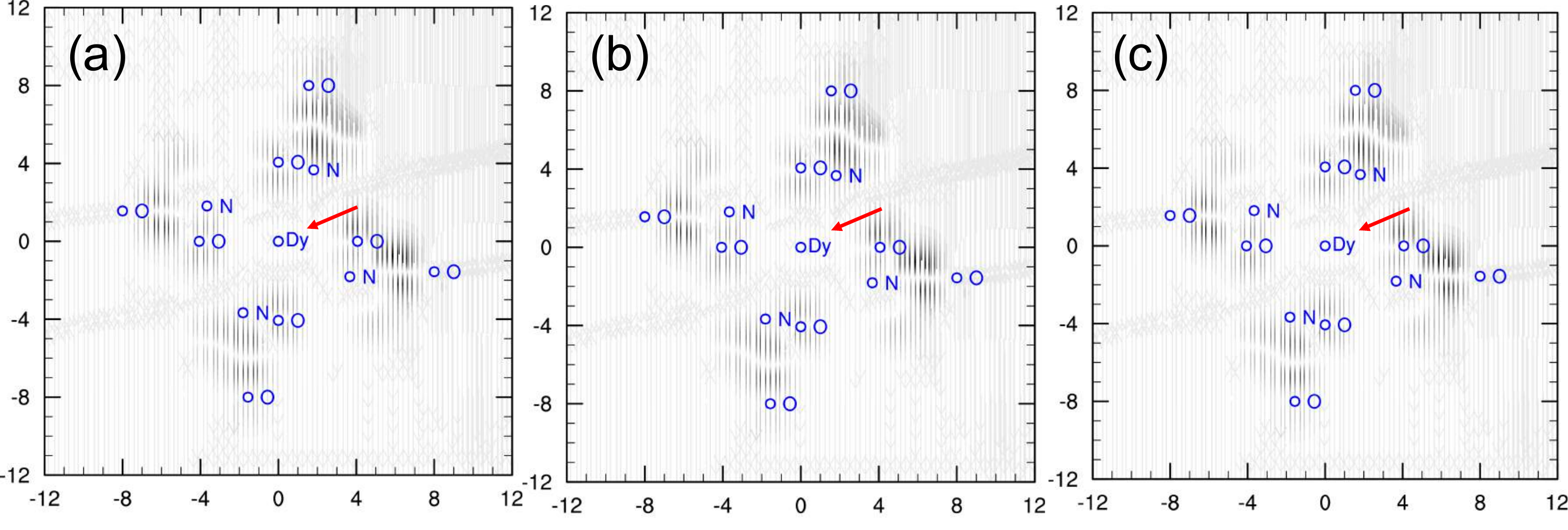}
	\caption{Comparison of the current density $\textbf{j}$\ obtained from DMRG-CI calculations with (a) CAS(9,14), (b) CAS(27,32), and (c) CAS(27,58) active orbital spaces. Visualization of $\textbf{j}$\ in the molecular plane containing the Dy(III) center in the (0,0,0) origin as indicated by a red arrow. Blue circles indicate the position of atomic centers where for better readability only the Dy(III)\ center as well as the oxygen and nitrogen centers of the chelating (\mmod)$^{4-}$ ligand are shown. Line intensity is proportional to the norm of \textbf{j}.}
	\label{fig:j-comp}	
\end{figure}
Moreover, these conclusions are corroborated 
by the streamline plots of the current density \textbf{j}\ in the molecular plane containing the Dy(III) center in the (0,0,0) origin shown in Figure \ref{fig:j-comp}\ as a function of increasing active orbital 
space size going from the smallest, CAS(9,14) (panel (a) in Figure \ref{fig:j-comp})\ to the largest one, CAS(27,58) (panel (c) in Figure \ref{fig:j-comp}). 
The line intensities of the charge current density streamline plots are in close agreement both qualitatively and quantitatively for all three active spaces under consideration. Interestingly, the current density field appears to have sources and drains --- the latter corresponding to a fading of the streamline intensity --- that are in regions of space considerably 
distant from the central Dy(III) ion of the \dyc\ complex. 
\begin{figure}[tbh]
\centering
	\includegraphics[scale=0.515]{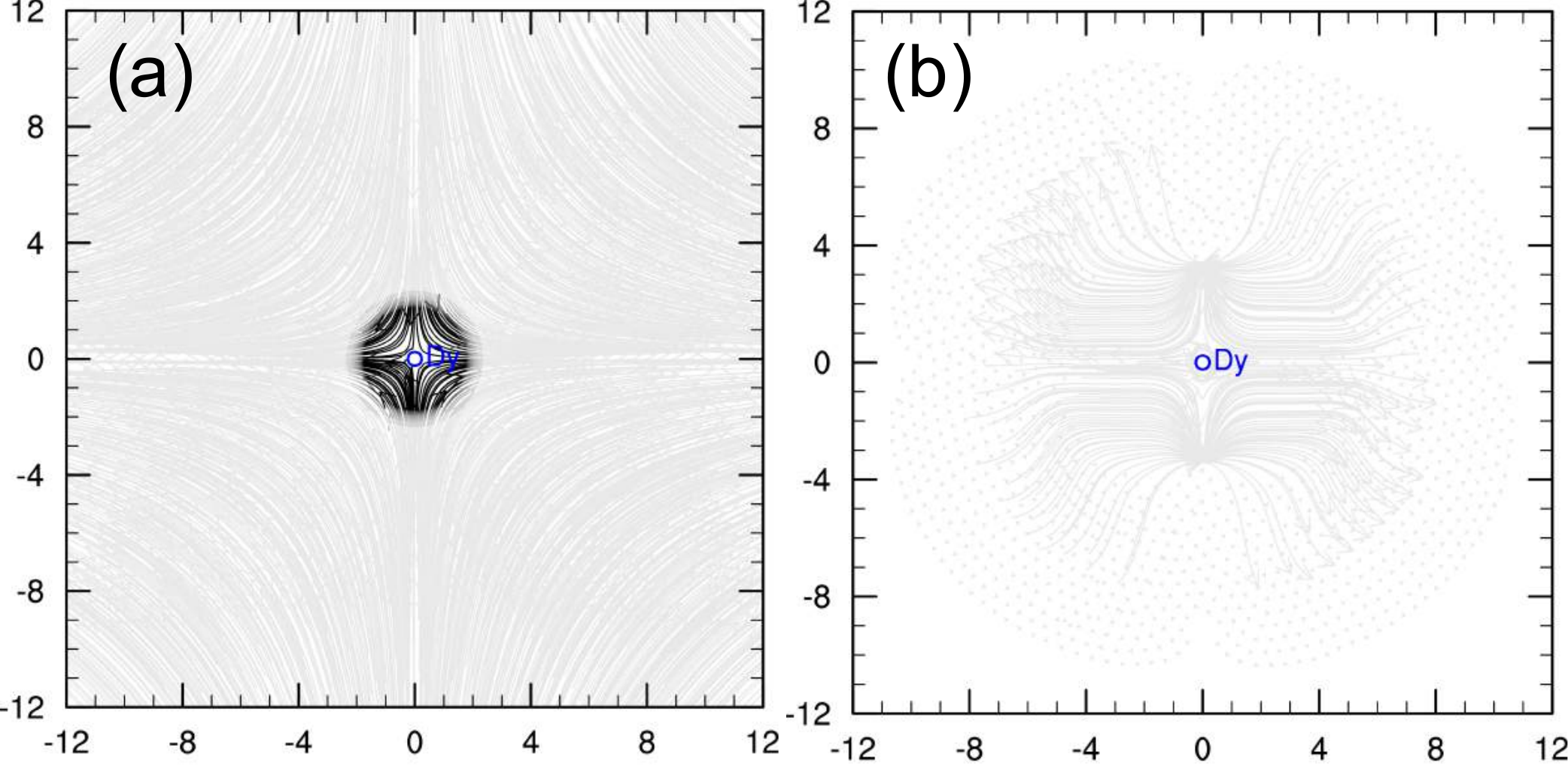}
	\caption{Streamline plots of the current density $\textbf{j}$\ (panel a) and magnetization $\textbf{m}$\ (panel b) in the plane comprising  
			the Dy(III) center at the (0,0,0) origin. Data was obtained from a DMRG-CI/CAS(9,14) calculation of the bare Dy$^{3+}$\ ion. Line intensities are proportional to the norm of \textbf{j}\ and \textbf{m}, respectively.}
	\label{fig:j-m-ion}	
\end{figure}

\begin{figure}[tbh]
\centering
	\includegraphics[scale=0.515]{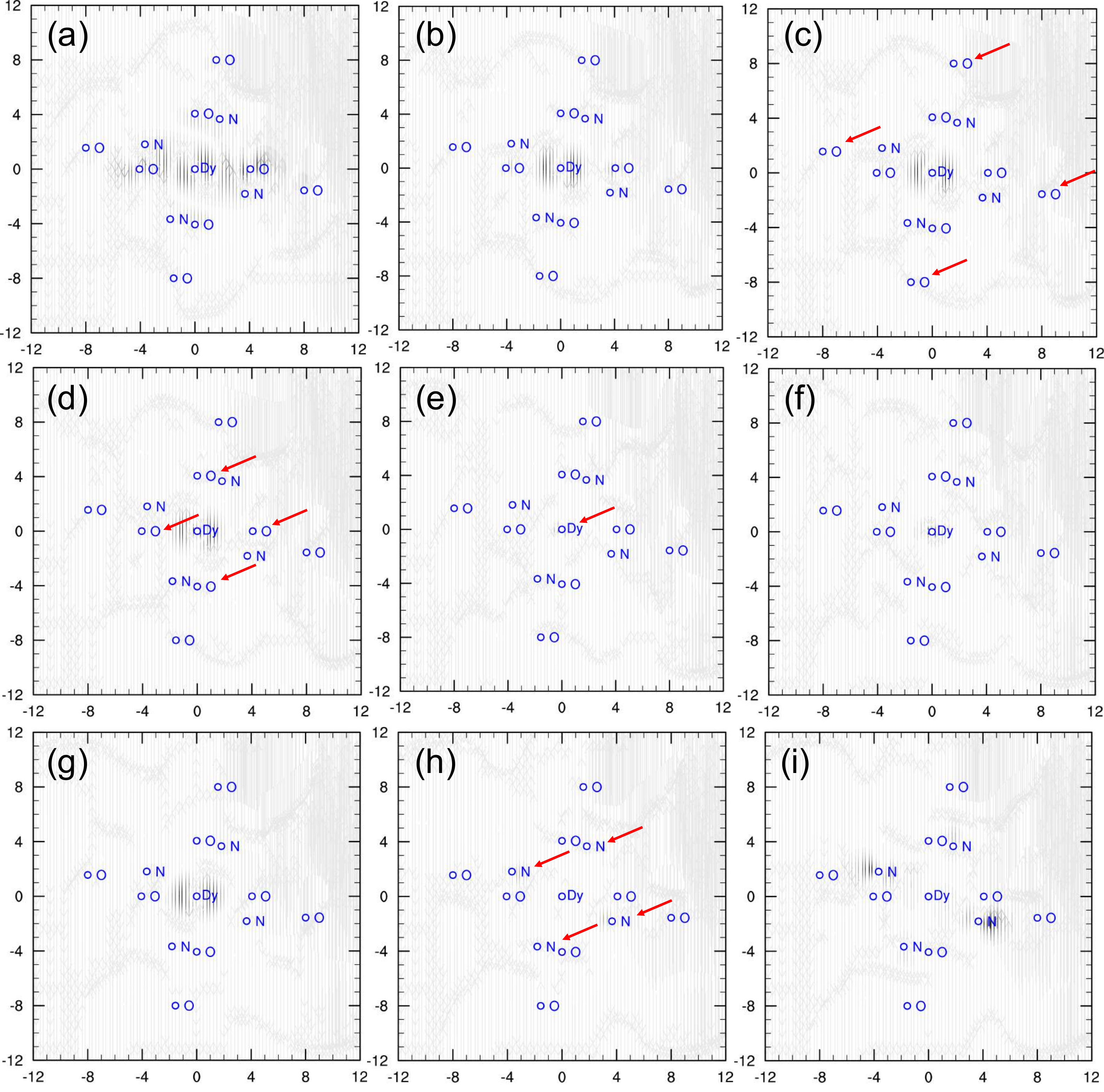}
	\caption{Magnetization $\textbf{m}$\ obtained from a DMRG-CI/CAS(27,32) calculation and 
		visualized in different molecular planes from above (subfigures (a)-(d)) to below (subfigures (f)-(i)) the plane in $z$-direction which contains the Dy(III) center (subfigure (e)) in the (0,0,$z=0$) origin. Blue circles indicate the position of atomic centers. 
				For better readability, only the Dy(III)\ center as well as the O and N centers of the chelating (\mmod)$^{4-}$ ligand are shown. 
				Red arrows indicate an integration plane containing the highlighted atomic center(s). Line intensities are proportional to the norm of \textbf{m}.}
	\label{fig:m}
\end{figure}

Turning to streamline plots for 
the bare Dy$^{3+}$\ ion in Figure \ref{fig:j-m-ion}, 
this picture changes significantly for both the magnetization 
(right-hand side of Figure \ref{fig:j-m-ion}) and, 
particularly the current density field (left-hand side of Figure \ref{fig:j-m-ion}). We find in either case sizable streamline intensities spatially close 
to the core and valence region of the atomic center as could 
be expected for an open-shell $4f^9$\ ion. 
Although putting the Dy(III) central ion into the crystal field of a chelating ligand such as \mmod\ did not seem to have a quantifiable influence on the local electronic ground-state structure of the resulting molecular 
\dyc\ complex (cf.~Figure \ref{fig:entanglement_dycomplex_1}), 
ligand-field effects clearly induce a redistribution of the current density and magnetization, causing a density shift towards the ligand centers. 
\begin{figure}[h!]
\centering
	\includegraphics[scale=0.515]{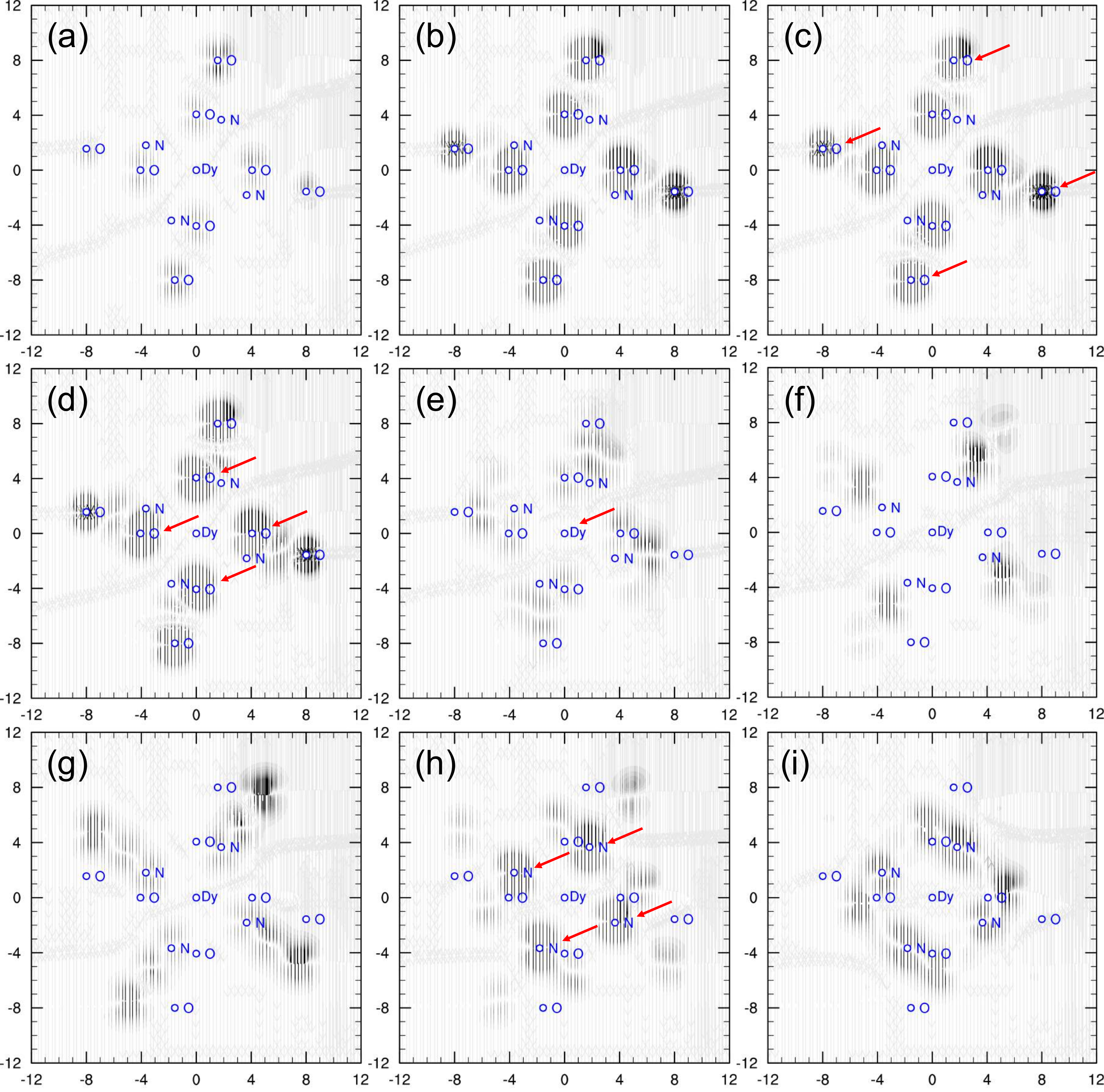}
	\caption{Current density $\textbf{j}$\ obtained from a DMRG-CI/CAS(27,32) calculation and 
		visualized in different molecular planes from above (subfigures (a)-(d)) to below (subfigures (f)-(i)) the plane in $z$-direction which contains the Dy(III) center (subfigure (e)) in the (0,0,$z=0$) origin. Blue circles indicate the position of atomic centers. 
		For better readability, only the Dy(III)\ center as well as the O and N centers of the chelating (\mmod)$^{4-}$ ligand are shown. 
		Red arrows indicate an integration plane containing the highlighted atomic center(s). Line intensities are proportional to the norm of \textbf{j}.}
	\label{fig:j}
\end{figure} 
The latter is illustrated in Figures \ref{fig:j} and \ref{fig:m}\ for different molecular planes below and above the (0,0,$0$) plane comprising the Dy(III) center. 
Considering first the current density fields in Figure \ref{fig:j}, 
we find stronger currents in particular in the planes containing the oxygen centers of the carboxylate groups of the chelating \mmod\ ligand 
compared to weaker ones within the plane spanned by the N atoms of the tetraaza-cyclododecane ring. By contrast, the situation is less clear for the magnetization, the spin-density analogue in a four-component relativistic framework, as can be seen from Figure \ref{fig:m}. Notable line intensities of the magnetization streamline plots can be found above and below the Dy(III)-containing plane but they do not seem to coincide with particular atomic centers of the \mmod\ ligand. 

In summary, by comparison to the bare Dy$^{3+}$\ ion our findings for the current density and magnetization of the \dyc\ complex 
substantiate the role of lanthanide chelating tags such 
as the Dy(III)-M8 complex presented first in Ref.~\citenum{haus09}\ 
as a versatile paramagnetic label for the NMR spectroscopy of 
biomolecules (in solution). 
For example, lanthanide chelating tags play a particular role in the determination of structural parameters of a tagged biomolecule 
by virtue of (paramagnetic) NMR spectroscopic techniques \cite{blea72}. 
They give rise to (significant) pseudocontact shifts and sizable Fermi contact shifts \cite{pigu03}, both of which are together commonly referred to as hyperfine paramagnetic lanthanide induced shifts.  
These shifts originate from a large magnetic anisotropy of the paramagnetic center (pseudocontact shift) and from a 
non-zero spin-density at the probed (light-atom) nucleus caused by spin polarization through the spin-density 
of the paramagnetic central ion (Fermi contact shift) \cite{kurl70}, respectively. Concluding from the above findings for 
the current density and magnetization distributions for the modified \dyc\ complex\ both could be expected to 
be large in magnitude at nuclear centers distant from the paramagnetic Dy(III) center. These results are in line with the experimentally determined pseudocontact shifts based on paramagnetic NMR measurements 
employing the unmodified Dy(III)-M8 complex in Ref.~\citenum{haus09}.

\section{Conclusions and Outlook}\label{sec:conclusions}
In this work we presented the formulation and efficient implementation of a genuine relativistic DMRG approach that exploits a matrix-product 
representation of the Hamiltonian as well as the wave function. To this end, 
we introduced a Kramers-paired spinor basis as well as double group symmetry into an existing nonrelativistic MPS/MPO-based DMRG algorithm. 
To demonstrate the suitability of our new relativistic DMRG model 
towards an unbiased exploration of the chemistry and properties of heavy element complexes, 
we considered two sample molecular systems. 
As a first example, we revisited the electron correlation problem 
in the electronic $\Omega=0^+$\ ground state of the diatomic molecule TlH that became a standard test system for new relativistic correlated 
electronic structure electron approaches. With our relativistic DMRG model 
we were able to reach near full-CI accuracy for the absolute energy of the electronic ground state close to the equilibrium structure by considering 
an active orbital space as large as comprising 14 electrons in 94 spinors. 
Yet, to determine spectroscopic properties to high accuracy requires 
firstly to take into account dynamical electron correlation effects which 
can only be described to a certain extent in a DMRG model. Hence, we 
currently pursue the combination of our relativistic DMRG approach with (i) 
multireference perturbation theories such as CASPT2 
\cite{abem06,kimi14,shio15a} and NEVPT2 \cite{shio15a} as well as 
with (ii) a range-separated DFT \textit{ansatz}\ \cite{hede15b} that showed 
promising potential in a nonrelativistic framework. 
Secondly, orbital relaxation in the presence of SO coupling and explicit 
consideration of electron correlation is another crucial aspects striving for a 
high-accuracy computational model. In a recent work \cite{mayi17b}, we 
demonstrated the robustness and efficiency of a second-order orbital 
optimization algorithm based on an MPS reference wave function. The 
formulation and implementation of its equivalent approach for a relativistic 
DMRG model is under way in our laboratory\ which will complement our locally 
available quasi-second-order DMRG-SCF implementation in \texttt{Bagel} 
\cite{bagel}\ that was originally presented for a CI reference wave function 
\cite{bate15}. 

Magnetic properties offer a unique insight into the spectroscopy of 
open-shell molecules such as our second sample molecule, an $f^9$\ \dyc\ complex. In this work, we investigated ligand-field effects on the magnetization and current density originating from the central Dy(III)\ ion by virtue of a real-space visualization of the density fields. In a graphical representation, 
we could illustrate an efficient spatial distribution of current density and magnetization originating from the paramagnetic Dy(III) center in direction to 
the ligand centers. The latter is a prerequisite for molecular complexes such as \dyc\ to serve as paramagnetic tags for biomolecules, making those available to paramagnetic NMR measurements. 
Consequently, an obvious extension of our expectation-value based property calculation approach for MPS wave functions 
will be the determination of first-order magnetic properties such as 
EPR g-tensors and hyperfine A-tensors. 
In a relativistic (four-component) framework, their calculation can be formulated as simple expectation values of well-defined one-electron operators \cite{vadm13}. Work in this direction is currently in progress.

\section*{Acknowledgments}

This work was supported by the Schweizerischer Nationalfonds (SNF project 200020\_169120). We particularly thank Prof. Markus Reiher, ETH Z\"urich, for insightful 
discussions and continuous support throughout this project. We are grateful to Dr.~Radovan Bast, Troms{\o}, 
for providing assistance with the visualization of the magnetization as well as the current densities. SK would like to thank Andrea Muolo for fruitful discussions and assistance with the 
interface of \qcm\ to the \texttt{BAGEL}\ software.


\newcommand{\Aa}[0]{Aa}

\end{document}